%% file: main.tex
\documentclass[journal]{IEEEtran}
\IEEEoverridecommandlockouts

\usepackage{cite}
\usepackage{amsmath,amssymb,amsfonts}
\usepackage{algorithm}
\usepackage{algorithmic}
\usepackage{diagbox}
\usepackage{subcaption}
\usepackage{adjustbox}
\usepackage{colortbl}
\usepackage{graphicx} 
\usepackage{subcaption}

\usepackage{stfloats}  
\usepackage{caption}    

\usepackage{footmisc}

\usepackage{multirow} 
\usepackage{pifont}
\usepackage{changes}
\definechangesauthor[name={Tongyu Song}, color=orange]{AF}

\usepackage{makecell}
\usepackage{graphicx}
\usepackage{hyperref}
\usepackage{textcomp}
\usepackage{xcolor}
\def\BibTeX{{\rm B\kern-.05em{\sc i\kern-.025em b}\kern-.08em
    T\kern-.1667em\lower.7ex\hbox{E}\kern-.125emX}}
\begin{document}

\title{Efficient Resource Allocation Framework for LoRaWAN Network via Online Learning}
\author{Ruiqi~Wang,~\IEEEmembership{Student Member,~IEEE,}
        Wenjun~Li,
        Jing~Ren,~\IEEEmembership{Member,~IEEE,}
        Tongyu~Song,~\IEEEmembership{Member,~IEEE,} 
        Xiong~Wang,~\IEEEmembership{Member,~IEEE,}
        Sheng~Wang,~\IEEEmembership{Member,~IEEE,}
        and~Shizhong~Xu,~\IEEEmembership{Member,~IEEE}
\thanks{Corresponding authors: Jing Ren and Tongyu Song.}
\thanks{Ruiqi Wang, Wenjun Li, Jing Ren, Xiong Wang, Sheng Wang, and Shizhong Xu are with the School of Information and Communication Engineering, University of Electronic Science and Technology of China, Chengdu 611731, China (email: 202421011308@std.uestc.eud.cn; 202522011107@std.uestc.edu.cn; renjing@uestc.edu.cn; wangxiong@uestc.edu.cn; wsh\_keylab@uestc.edu.cn; xsz@uestc.edu.cn). Tongyu Song is with the Research Institute of Intelligent Networks, Zhejiang Lab, Hangzhou 311121, China (email: tongyusong@ieee.org).}}


\maketitle
 
\begin{abstract}
The deployment of large-scale LoRaWAN networks requires jointly optimizing conflicting metrics like Packet Delivery Ratio (PDR) and Energy Efficiency (EE) by dynamically allocating transmission parameters, including Carrier Frequency, Spreading Factor, and Transmission Power. Existing algorithms often ignore the complexity of multi-objective dynamic adaptation to oversimplify this challenge, focusing on a single metric or lacking the adaptability needed for dynamic channel environments, leading to suboptimal performance. To address this, we propose two online learning-based resource allocation frameworks that intelligently navigate the PDR-EE trade-off. Our foundational proposal, D-LoRa, is a fully distributed framework that models the problem as a Combinatorial Multi-Armed Bandit. By decomposing the joint parameter selection and employing specialized, disaggregated reward functions, D-LoRa dramatically reduces learning complexity and enables nodes to autonomously adapt to network dynamics. To further enhance performance in LoRaWAN networks, we introduce CD-LoRa, a hybrid framework that integrates a lightweight, centralized initialization phase to perform a one-time, quasi-optimal channel assignment and action space pruning, thereby accelerating subsequent distributed learning. Extensive simulations and real-world field experiments demonstrate the superiority of our frameworks, showing that D-LoRa excels in non-stationary environments while CD-LoRa achieves the fastest convergence. In physical deployments, our algorithms outperform state-of-the-art baselines, improving PDR by up to 10.8\% and EE by 26.1\%, confirming their practical effectiveness for scalable and efficient LoRaWAN networks.
\end{abstract}


\input{sections/1-intro}

\input{sections/2-relatedwork}
\input{sections/3-model}

\input{sections/4-method}

\input{sections/5-simulation}
\input{sections/6-field}
\input{sections/7-conclusion}

\input{sections/0-bib}
\end{document}

%% file: sections/1-intro.tex
\section{Introduction}
The rapid growth of the Internet of Things (IoT) necessitates robust and scalable connectivity solutions for a massive number of devices. Long Range Wide Area Network (LoRaWAN) \cite{LPWAN,LoRaWAN} deployed in a star topology has been proposed as a pivotal IoT solution. It employs LoRa (Long Range) as its physical radio frequency modulation method, which is known for its long-range communication capabilities, low power consumption, and cost-effectiveness \cite{survey1, survey2, survey3, survey4}. These features make LoRaWAN well-suited for a diverse range of applications, including smart agriculture \cite{survey3,survey4}, urban infrastructure management \cite{survey1},\cite{survey3}, and industrial applications \cite{survey1}, \cite{survey4}.

Effective resource allocation in the LoRaWAN network is crucial for enhancing two key performance metrics: Packet Delivery Ratio (PDR) for network reliability and Energy Efficiency (EE) for extended device operational lifetime. 
Improperly allocating transmission parameters (i.e., radio resources), including Carrier Frequency (CF), Spreading Factor (SF), and Transmission Power (TP), increases packet losses and causes excessive energy consumption, which severely degrades network performance.  
Therefore, developing an intelligent resource allocation strategy for CF, SF, and TP is essential for enhancing LoRaWAN network performance by mitigating interference and conserving energy simultaneously \cite{survey3,survey4}.
  
However, optimizing the transmission parameters in a LoRaWAN network is a intricate problem due to several inherent complexities. 
Firstly, the joint allocation of CF, SF, and TP to a large number of devices constitutes a combinatorial optimization problem, which is known to be NP-hard \cite{nphard1,nphard2}. 
Secondly, accurately modeling the physical channel behavior, particularly the imperfect quasi-orthogonality between different SFs that leads to inter-SF interference, remains a significant challenge \cite{Orthogonality}.
Most critically, the intricate relationships among these parameters create fundamental trade-offs between PDR and EE. 
For instance, assigning a lower SF reduces a packet's Time on Air (ToA), thereby conserving energy. 
However, this simultaneously shrinks the communication range, potentially degrading the PDR for distant devices. 
Conversely, increasing the TP can boost the PDR by overcoming poor channel conditions, but this often leads to excessive energy consumption, especially for devices near the gateway, thus diminishing the overall EE.
This intricate coupling relationship means that simplistic or greedy allocation strategies often improve one metric at the expense of the other, highlighting the necessity for a sophisticated approach that can navigate these complex interdependencies to achieve a balanced network performance.
Due to the complexity of the problem, existing work can be computationally burdensome or oversimplify the problem by focusing on single parameter/metric. Besides, they often ignore the adaptability needed for dynamic environments, leading to suboptimal performance.

To tackle this complex optimization challenge, we introduce two novel online learning resource allocation frameworks, progressing from a fully distributed to a hybrid framework, both designed to enhance EE while ensuring a high PDR.
Our foundational proposal is D-LoRa \cite{D-LoRa}, a fully distributed framework based on the Combinatorial Multi-Armed Bandit (CMAB) framework \cite{CMAB}. 
The motivation behind this distributed framework is to empower each end-device with autonomy, eliminating the need for constant, costly communication with a central server and dynamically allocating parameters to adapt to the changing channels. 
In D-LoRa, an agent on each device independently learns a policy to select the optimal CF, SF, and TP, aiming to improve EE while maintaining high PDR.
However, a purely distributed learning process can suffer from slow convergence and high exploration overhead, especially in a large action space encompassing all parameter combinations. 
To address this practical limitation, we propose an hybrid framework, CD-LoRa (Centralized-assisted D-LoRa). 
This approach is designed to accelerate the learning process and reduce on-device computational burden. 
It initiates with a lightweight, centralized procedure at the gateway, termed Channel Allocation and Action Space Initialization (CAASI). 
Leveraging easily accessible RSSI data, CAASI performs a one-time, coarse-grained optimization: it assigns a near-optimal static channel to each device and prunes its SF action space by eliminating choices that are clearly unsuitable. 
Following this initialization, a distributed CMAB agent, operating within this significantly reduced and better-informed action space, can perform online learning to fine-tune the remaining SF and TP parameters much more efficiently.
The main contributions of this paper are as follows:
\begin{itemize}
\item \textbf{A Comprehensive Analytical Model and Problem Formulation:} We develop a detailed analytical model for LoRaWAN networks that captures key physical layer characteristics, including path loss, inter-SF interference arising from imperfect quasi-orthogonality, and packet collisions. 
Based on this model, we formulate the resource allocation challenge as a formal optimization problem aimed at co-optimizing network PDR and EE.
\item \textbf{A Fully Distributed, Bandit-based Resource Allocation Framework (D-LoRa):} We design D-LoRa, a novel distributed framework that reframes the complex parameter selection task as a CMAB problem. 
By decomposing the joint parameter selection into independent base arms (CF, SF, TP), D-LoRa significantly reduces the learning complexity.
Furthermore, it employs specially crafted reward functions that explicitly navigates the intrinsic trade-off between PDR and EE, enabling autonomous on-device optimizations.
\item \textbf{A Hybrid Framework for Accelerated Learning (CD-LoRa):} To overcome the potential for slow convergence in purely distributed systems, we propose CD-LoRa, a hybrid framework. 
It introduces a lightweight, gateway-assisted initialization phase (CAASI) that performs a one-time, coarse-grained optimization to assign static channels and prune the action space for each device. 
This synergy between centralized preprocessing and distributed online learning allows CD-LoRa to achieve faster convergence and superior performance with reduced on-device overhead.
\item \textbf{Experimental Evaluation:} We validate the proposed D-LoRa and CD-LoRa frameworks through extensive simulations and real-world field experiments. The results demonstrate their superiority over four baseline algorithms in achieving a more effective PDR-EE trade-off, under both stationary, nonstationary scenarios.
\end{itemize}

The remainder of this article is organized as follows. Section \ref{relatedwork} gives a brief introduction to the related work about resource allocation in the LoRaWAN network. In Section \ref{model}, we establish the system model and formulate the optimization problem. In Section \ref{method}, we give a detailed introduction to our frameworks. Section \ref{simulation} and Section \ref{field} present the results and analysis of the simulation experiments and field experiments, respectively. Finally, we give a conclusion about this work in Section \ref{cons}.

%% file: sections/2-relatedwork.tex
\section{Related Work}
\label{relatedwork}
Resource allocation algorithms in LoRaWAN networks are implemented through three primary frameworks: centralized, distributed, and hybrid.

\textbf{Centralized Framework:} With a centralized framework, the central coordinator (e.g. Gateway or Network Sever) is responsible for global optimization. 
The central coordinator gathers network-wide information, such as link quality indicators from all end-devices, to compute and disseminate optimal transmission parameters through downlink messages \cite{central1,central2,central3,central4,central5,central6,central7,central8,central9}. 
A representative example of these algorithms is the Adaptive Data Rate (ADR) mechanism \cite{central1}, defined in the LoRaWAN protocol. 
ADR aims to conserve energy while preserving link reliability by dynamically adjusting the SF and TP for each end-device based on its historical signal quality.
However, the conventional ADR mechanism has been shown to exhibit several limitations, which has motivated the proposal of numerous improvement algorithms.
For instance, Li et al. \cite{central5} argued that ADR's simple reliance on Signal-to-Noise Ratio (SNR) thresholds leads to suboptimal energy efficiency and proposed DyLoRa, which utilizes a more sophisticated symbol error rate-energy efficiency model. 
Zhao et al. \cite{central4} identified that both ADR and DyLoRa ignore the issue of fairness and propose EF-LoRa as a solution.
EF-LoRa allocates parameters by maximizing the minimum energy efficiency among all devices, thereby enforcing max-min fairness. 
More recently, to enhance network lifetime, Yang et al. \cite{central9} developed RALoRa, which combines block-level rateless coding with joint network-wide optimization and real-time path-loss prediction.

Despite their potential for achieving global optimality, centralized approaches suffer from significant practical drawbacks. 
First, the requisite exchange of control messages for state collection and parameter dissemination imposes substantial signaling overhead, which can be prohibitive in downlink-constrained LoRaWAN networks. 
Second, the computational and storage burden on the central server grows with network size, posing a critical scalability challenge. 
Finally, the centralized architecture introduces a single point of failure, where a server malfunction can compromise the entire network's adaptation capability. These limitations motivate the exploration of distributed and hybrid frameworks.

\textbf{Distributed Framework:}  In response to the drawbacks of centralized control, the distributed framework empowers end-devices to autonomously adapt their transmission parameters. 
In this framework, each device leverages local information to make decisions \cite{distributed1,distributed2,distributed3,distributed4,distributed5,distributed6,distributed7,distributed8,distributed9}, thereby obviating the need for a central coordinator.
The inherent uncertainty in distributed online decision-making has spurred the application of machine learning techniques. 
Among these, the Multi-Armed Bandit (MAB) framework has emerged as a particularly suitable model, as it naturally formulates the problem of an agent learning the best action (i.e., parameter combination) under dynamic environment through repeated trials and feedback.
Early work by Bonnefoi et al. \cite{distributed1} pioneered the use of MAB for channel selection, enabling IoT devices to independently learn optimal channel access aiming to improve successful transmission rate. 
Building on this, Azari et al. \cite{distributed2} extended the MAB framework to incorporate energy consumption, allowing devices to learn the most energy-efficient parameters without network-level signaling. 
Subsequent efforts focused specifically on SF optimization in LoRaWAN network, with proposals like LoRa-MAB and its computationally lighter variant, EXP3.S \cite{distributed3}.
More recent work, such as MIX-MAB \cite{distributed4}, has addressed the joint optimization of CF, SF, and TP by combining Successive Elimination (SE) and EXP3, which successfully improves convergence speed and PDR. Beyond the MAB framework, other learning-based approaches have also been explored. 
Hong et al. \cite{distributed6} employed a Temporal Difference (TD) reinforcement learning algorithm for SF allocation, while Scarvaglieri et al. \cite{distributed9} proposed a cooperative multi-agent Q-learning algorithm for distributed resource management.

Despite these advances, limitations remain in the literature on distributed algorithms. 
First, many proposed algorithms concentrate on optimizing a single parameter (e.g., only SF or CF) or target a singular performance metric (e.g., only PDR), rather than handling the complex coupling between parameters and performance metrics. 
And empirical validation through real-world deployments remains scarce. 
These highlight the need for a distributed scheme that can allocate multiple transmission parameters while handling the trade-offs among different performance metrics, with its effectiveness verified in a real-world setting.

\textbf{Hybrid Framework:} The hybrid framework chooses to combine the benefits of centralized and distributed frameworks by leveraging a central entity for global network oversight while preserving the autonomy of end-devices for local adaptation \cite{hybrid1,hybrid2,hybrid3}. 
For instance, Reynders et al. \cite{hybrid1} proposed RS-LoRa, a two-stage scheduling algorithm that combines gateway-initiated broadcasts with device-level self-scheduling to improve network scalability. 
Other works have explored more intricate integrations of learning and optimization. 
CORA \cite{hybrid2}, for example, employs a sophisticated two-step algorithm combining multi-agent reinforcement learning with convex optimization to jointly manage CF, SF, and TP.
However, its significant computational complexity may be a barrier to large-scale, practical deployments.
More recently, research has focused on using centralized information to enhance distributed learning. 
Zhang et al. \cite{hybrid3} proposed LI-WEX, which augments MAB-based techniques by incorporating link prior knowledge to better balance the PDR and EE trade-off. 
While simulation results have demonstrated the superiority of the LI-WEX algorithm, the algorithm's link prior knowledge relies solely on the mathematical model that may not fully capture the dynamics of actual transmission environments. Besides, its real-world performance remains to be validated, making the practical efficacy of the algorithm uncertain.

Collectively, these hybrid strategies underscore a promising research direction. 
Existing algorithms can be computationally burdensome or lack the empirical validation necessary to prove their practicality. 
This motivates the development of a novel hybrid framework that strategically leverages easily accessible global information (such as RSSI) to reduce the operational complexity of distributed algorithms, thereby accelerating convergence and improving efficiency in a manner that is verifiable through physical implementation.

%% file: sections/3-model.tex
\section{System Model and Problem Formulation}
\label{model}
In this section, we first introduce the system modeling and then formulate the constrained optimization problem towards optimizing PDR and EE.
\subsection{Network Model}
In this paper, we consider a star topology LoRaWAN network with a central gateway GW  and a set $\mathcal{N}=\{n_{1},n_{2},...,n_{N}\}$ of $N=|\mathcal{N}|$ nodes randomly distributed in the area. The nodes collect data from the surrounding environment and send the encoded packets to the GW. The set $\mathcal{P}=\{p_{1},p_{2},...,p_{P}\}$ denotes the $P=|\mathcal{P}|$ packets sent by these nodes. The definitions are shown below:

\textit{Definition 1 (Gateway)}: The central gateway is denoted as $\text{GW} \doteq\left \langle x^{\text{g}},y^{\text{g}},\mathcal{P}^{\text{gr}}\right \rangle$, where $(x^{\text{g}},y^{\text{g}})$ is the two-dimensional coordinate of the gateway and $\mathcal{P}^{\text{gr}}$ is the set of the packets the gateway received.

\textit{Definition 2 (Node)}: A node $i$ is denoted as $n_{i}\doteq \left \langle x_{i},y_{i},\mathcal{P}^{\text{ns}}_{i},\mathcal{P}^{\text{nr}}_{i},\mathcal{P}^{\text{nl}}_{i},\mathcal{LR} \right \rangle$, where $(x_{i},y_{i})$ is the two-dimensional coordinate of the node. $\mathcal{P}^{\text{ns}}_{i}$ is the set of the packets sent by node $i$, $\mathcal{P}^{\text{nr}}_{i}$ is the set of the packets sent by node $i$ that are successfully received and $\mathcal{P}^{\text{nl}}_{i}$ is the set of the packets sent by node $i$ that are lost. $\mathcal{LR} \doteq \{\mathcal{SF},\mathcal{CF},\mathcal{TP}\}$ is the set of available LoRa parameters.

\textit{Definition 3 (Packet)}: A packet $j$ is denoted as $p_{j}\doteq\left \langle \text{PS}_{j},\text{id}(j),\mathcal{LP}_{j},E_{j}\right \rangle$, where $\text{PS}_{j}$ is the payload size of the packet and $\text{id}(j)$ is the identity of the node that generates packet $j$. 
$\mathcal{LP}_{j}=\{\text{SF}_{j},\text{CF}_{j},\text{TP}_{j}\}$ is the LoRa parameters selected by $n_{\text{id}(j)}$ to configure the packet. $E_{j}$ is the energy consumption of $n_{\text{id}(j)}$ sending $p_{j}$.

\subsection{Packet Collision Model}
\label{CollisionModel}
Authors in \cite{LoRaSim} proposed the definitions of four kinds of packet collisions: CF, SF, power, and timing. The binary variable $C_{j}$ indicating whether $p_{j}$ collides during transmission can be expressed as \cite{LoRaSim}:
\begin{equation}
   C_{j}=C^{\text{time}}_{j}\wedge C^{\text{SF}}_{j}\wedge C^{\text{CF}}_{j}\wedge C^{\text{pwr}}_{j},
    \label{Collision}
\end{equation}
where $C^{\text{time}}_{j}$, $C^{\text{SF}}_{j}$, $C^{\text{CF}}_{j}$, and $C^{\text{pwr}}_{j}$ are binary variables indicating whether $p_{j}$ has time overlap, SF collision, CF collision, and capture effect with other packets. $C_{j}=1$ means that $p_{j}$ is lost due to collision during transmission.

\subsection{Packet Propagation Model}
Given the application scenarios in densely populated areas, we employ a radio transmission model based on the Log-Distance Path Loss Model. The path loss $L^{\text{pl}}_{j}(d)$ of $p_{j}$ when transmitted to the gateway can be expressed as:
\begin{equation}
   L^{\text{pl}}_{j}(d)=\overline{L^{\text{pl}}}(d_{0})+10\gamma \text{log}(\frac{d}{d_{0}})+X_{\delta},
    \label{LoS}
\end{equation}
where $\overline{L^{\text{pl}}}(d_{0})$ is the average path loss with the reference distance is $d_{0}$, $d=\sqrt{(x_{\text{id}(j)}-x^{\text{g}})^{2}+(y_{\text{id}(j)}-y^{\text{g}})^{2}}$ is the Euclidean distance between $n_{\text{id}(j)}$ and GW, $\gamma$ is the path loss factor, $X_{\delta_{1}}\sim N(0,\delta_{1}^{2})$ is the normal distribution considering the shadowing effect. Assuming that the effects of other gains and losses in the propagation process are zero, the RSSI of $p_{j}$ at GW can be expressed as:
\begin{equation}
   \text{RSSI}_{j}=\text{TP}_{j}-\overline{L^{\text{pl}}}(d_{0})-10\gamma \text{log}(\frac{d}{d_{0}})-X_{\delta}.
    \label{RSSI}
\end{equation}
$p_{j}$ can only be successfully decoded by GW if $\text{RSSI}_{j}$ is not less than its Receiver Sensitivity ($\text{RS}$). $\text{RS}_{j}$ is determined by the $\text{\text{SF}}_{j}$ and bandwidth (BW) of $p_{j}$, which can be expressed as:
\begin{equation}
   \text{RS}_{j}=-174+10\log_{10}(\text{BW})+\text{NF}+\text{SNR}_{j},
    \label{RS}
\end{equation}
where $\text{NF}$ is the Noise Figure which is a fixed value depending on the hardware, and $\text{SNR}_{j}$ depends only on the $\text{\text{SF}}_{j}$. The $\text{RSs}$ corresponding to different SF and BW combinations are shown in TABLE \ref{tab1}\cite{Datasheet}:
\begin{table}[htbp]
\caption{RSs in dBm for different SF and BW combinations}
\centering
\begin{tabular}{c|c|c|c|c|c|c}
\hline
\diagbox{\textbf{BW}}{\textbf{\text{SF}}}&\textbf{7}&\textbf{8}&\textbf{9}&\textbf{10}&\textbf{11}&\textbf{12} \\
\hline
125kHz&-123&-126&-129&-132&-133&-136\\
\hline
250kHz&-120&-123&-125&-128&-130&-133\\
\hline
500kHz&-116&-119&-122&-125&-128&-130\\
\hline
\end{tabular}
\label{tab1}
\end{table}\\
Given the quasi-orthogonality of SF, we incorporated the inter-SF interference into the Signal-to-Interference-plus-Noise Ratio (SINR) calculations for packets. The $\text{SINR}_{j}$ of $p_{j}$ can be expressed as:
\begin{equation}
   \text{SINR}_{j}=\frac{\text{RSSI}_{j}}{\sum_{k\ne j,\text{\text{SF}}_{k}\ne \text{\text{SF}}_{j},\text{\text{CF}}_{k}= \text{\text{CF}}_{j}}\text{RSSI}_{k}+\text{N}_{0}\text{W(}u,\delta^{2})},
    \label{RS}
\end{equation}
where $\sum_{k\ne j,\text{\text{SF}}_{k}\ne \text{\text{SF}}_{j},\text{\text{CF}}_{k}= \text{\text{CF}}_{j}}\text{RSSI}_{k}$ is interference from the packets having reception overlap with $p_{j}$ with the same CF and different SF. $\text{N}_{0}\text{W}(u,\delta_{2}^{2})$ is the Additative White Gaussian Noise (AWGN). $p_{j}$ can only be successfully decoded by GW when $\text{SINR}_{j}$ is not less than its SINR threshold ($\text{SINR}^{\text{thr}}_{j}$), which is determined by $\text{\text{SF}}_{j}$. The SINR thresholds for different SFs are shown in TABLE \ref{tab2}\cite{Datasheet}.
\begin{table}[htbp]
\caption{SINR thresholds in dB for different SF}
\centering
\begin{tabular}{c|c|c|c|c|c|c}
\hline
\textbf{\text{SF}}&\textbf{7}&\textbf{8}&\textbf{9}&\textbf{10}&\textbf{11}&\textbf{12} \\
\hline
$\textbf{\text{SINR}}^{\text{thr}}$&-7.5&-10&-12.5&-15&-17.5&-20\\
\hline
\end{tabular}
\label{tab2}
\end{table}

Based on the analysis presented above, $p_{j}$ can only be successfully received by GW if both $\text{RSSI}_{j}$ and $\text{SINR}_{j}$ are not less than the thresholds. $S_{j}$ is a binary variable that denotes whether $p_{j}$ is lost in the propagation process due to its weak signal strength, which can be expressed as:
\begin{equation}
   S_{j}=
   \begin{cases}
      0,& \text{if }\text{RSSI}_{j}\ge \text{RS}_{j}\text{ and }\text{SINR}_{j}\ge \text{SINR}^{\text{thr}}_{j},\\
      1,& \text{otherwise. } 
    \end{cases}
    \label{LossFlag}
\end{equation}
\subsection{Energy Consumption Model}
The symbol period of $p_{j}$ is determined by $\text{\text{SF}}_{j}$ and $\text{BW}$, which is calculated as $T^{\text{sym}}_{j}=\frac{2^{\text{\text{SF}}_{j}}}{\text{BW}}$.  The Time on Air (ToA) of $p_{j}$ can be expressed as:
\begin{equation}
    \begin{aligned}
        \text{ToA}_{j}&=T^{\text{pre}}_{j}+T^{\text{pay}}_{j},
    \end{aligned}
    \label{ToA}
\end{equation}
where $T^{\text{pre}}_{j}$ is the preamble duration and $T^{\text{pay}}_{j}$ is the payload duration, which can be expressed respectively as:
\begin{equation}
        T^{\text{pre}}_{j}=(n^{\text{pre}}+4.25)\cdot T^{\text{sym}}_{j}\text{, } T^{\text{pay}}_{j}=n^{\text{pay}}_{j}\cdot T^{\text{sym}}_{j}, 
    \label{Duration}
\end{equation}
where $n^{\text{pre}}$ is the preamble size of a LoRa packet, which is 8 symbols by default. $n^{\text{pay}}_{j}$ is the number of payload symbols of packet $j$, which can be calculated as:
\begin{equation}
    \begin{aligned}
        n_{j}^{\text{pay}}&=(8+\max (\left \lceil \frac{8\text{PS}_{j}-4\text{\text{SF}}_{j}+28+16\text{CRC}-20\text{H}}{4(\text{\text{SF}}_{j}-2\text{DE})}\right \rceil\\& (\text{CR}+4),0)).
    \end{aligned}
    \label{n_payload}
\end{equation}

In the default configuration of LoRaWAN, Cyclic Redundancy Check (CRC=1) is enabled, the header is enabled ($\text{H} = 0$), the LowDataRateOptimization is disabled ($\text{DE} = 0$), and the coding rate is set to $4/5$ ($\text{CR} = 1$). 
The energy consumed by $n_{\text{id}(j)}$ sending $p_{j}$ can be expressed as:
\begin{equation}
        E_{j}=\text{TP}_{j}\cdot \text{ToA}_{j}.
    \label{EnergyCon}
\end{equation}
\subsection{Optimization Problem Formulation}
In our work, we consider $\text{PDR}$, $\text{EE}$ as the network performance metrics. PDR is defined as the ratio of the number of packets successfully received by the GW to the total number of packets sent by the nodes in the network, which can be expressed as:   
\begin{equation}
        \text{PDR}=\frac{\lvert \mathcal{P}^{\text{gr}} \rvert}{\sum_{i=0}^{N} \lvert \mathcal{P}^{\text{ns}}_{i} \rvert}.
    \label{PDR}
\end{equation}
EE is defined as the amount of effective data that can be successfully transmitted per unit of energy consumed by the network in bits/mJ, which can be expressed as:
\begin{equation}        
    \text{EE}=\frac{\sum_{p_{i}\in\mathcal{P}^{\text{gr}}}\text{PS}_{i}}{\sum_{n_{j}\in\mathcal N}\sum_{p_{k}\in \mathcal{P}^{\text{ns}}_{j}}E_{k}},
    \label{EE}
\end{equation}
where the numerator represents the size of effective data received by the GW, and the denominator represents the total energy consumption of all nodes during the transmission process. 

Our objective is to improve the performance of the LoRaWAN network, which involves optimizing both PDR and EE rather than focusing on a single one. 
Hence, we have devised a utility function, comprising a weighted aggregation of the two metrics, to assess the network's performance quantitatively. The performance optimization problem is formulated as follows:
\begin{align}
    \text{(P1) }&\max \mathcal{U}=\alpha_1 \cdot \text{PDR} +\alpha_2 \cdot \text{EE} \label{problem} \\
    \text{s.t. } & C_{j}=0\land S_{j}=0, \forall p_{j}\in\mathcal{P}^{\text{gr}}\label{problemsub1}, \\
    & \text{SF}_{j}\in\mathcal{SF}, \text{CF}_{j}\in\mathcal{CF}, \text{TP}_{j}\in\mathcal{TP}, \forall p_{j}\in\mathcal{P} \label{problemsub2}, 
\end{align}
where $\mathcal{U}$ is the utility function, $\alpha_1$, $\alpha_2$ are the weight factors and $\alpha_1+\alpha_2=1$. Constraint (\ref{problemsub1}) means $p_{j}$ can only be received by GW when both its $\text{RSSI}_{j}$ and $\text{SINR}_{j}$ are no less the thresholds and there is no collision occurred during its transmission. Constraints (\ref{problemsub2}) limit the LoRa parameters that are available for nodes to configure for packet transmission.

%% file: sections/4-method.tex
\section{Algorithm Design}
\label{method}
This section develops our proposed resource allocation solutions, proceeding from a baseline model to progressively more sophisticated frameworks. We first establish a benchmark by formulating the problem as NaiveMAB, a straightforward MAB approach where each parameter configuration constitutes a single super arm. We then introduce D-LoRa, a fully distributed framework designed to overcome the scalability limitations of NaiveMAB by decomposing the action space into independent base arms. Building upon this, we present CD-LoRa, a hybrid framework that integrates a centralized initialization phase to accelerate learning convergence. The section concludes with a formal analysis of the algorithms' properties, focusing on their complexity and theoretical regret.
\subsection{Inspiration from the NaiveMAB}
In our proposed distributed framework, the parameter selection for packet transmission at each LoRa node is modeled as an MAB problem. Each node $i$ acts as an independent agent, learning the optimal transmission strategy based on its historical transmission results. 
In this formulation, a straightforward approach is to define each unique LoRa parameter configuration (CF, SF, TP) as a single super arm $\boldsymbol{a}\in \boldsymbol{\mathcal{A}}$ ($\mathcal{A}$ is the set of super arms). 
The agent's action for each packet transmission is thus the selection of one such super arm from the available set. Following each packet transmission, the agent observes the transmission result and receives an immediate numerical reward. 
This reward is then utilized to update the agent's expected reward of the selected super arm. 
To formalize this process, consider that at a given time step $t$, node $i$ selects a super arm $\boldsymbol{a}$ to configure the data packet $p_j$. After this action, it observes a reward, $r^t_{i}(\boldsymbol{a})$, which is formulated as:
\begin{equation}        
    r^t_{i}(\boldsymbol{a})=\mathbb{I}\{C_j=0\wedge S_j=0\},
    \label{naive_reward}
\end{equation}
where $\mathbb{I}\{C_j=0\wedge S_j=0\}$ denotes the indicator operator which is 1 if $p_j$ is successfully received and is 0 if $p_j$ is lost. 
Then, $n_i$ updates the expected reward $\overline{R}_i^t(\boldsymbol{a})$ for $\boldsymbol{a}$:
\begin{equation}        
    \overline{R}_i^t(\boldsymbol{a})=\overline{R}_i^{t-1}(\boldsymbol{a})+\frac{1}{T_i^t(\boldsymbol{a})}[r_i^t(\boldsymbol{a})-\overline{R}_i^{t-1}(\boldsymbol{a})],
    \label{naive_expected_reward}
\end{equation}
where $T_i^t(\boldsymbol{a})$ is the number of times $n_i$ chooses super arm $\boldsymbol{a}$ until $t$-th packet transmission. 

A core challenge in the MAB framework is the exploration-exploitation trade-off. 
When a super arm $\boldsymbol{a}$ has been selected only a few times (i.e., when $T_i^t(\boldsymbol{a})$ is small), its expected reward, $\overline{R}_i^t(\boldsymbol{a})$, is an unreliable estimator of its true reward \cite{EstimatedReward}. 
A purely greedy strategy, which exclusively selects the arm with the highest current expected reward, risks prematurely converging on a suboptimal arm that generated high rewards by chance. 
This phenomenon, known as premature exploitation, leads to significant performance degradation. 
To address this, we incorporate an exploration term into the decision-making process. 
This exploration term encourages the agent to select arms with high uncertainty, using an estimated reward for each super arm:
\begin{equation}           \widehat{R}_i^t(\boldsymbol{a})=\overline{R}_i^t(\boldsymbol{a})+c\cdot\sqrt{\frac{\log(t)}{2T_i^t(\boldsymbol{a})}}, \boldsymbol{a}\in\boldsymbol{A},
    \label{naive_estimated_reward}
\end{equation}
where $c$ is the weight factor, which is used to adjust the tendency towards exploration and exploitation. 
The exploration term is formulated based on the well-established Upper Confidence Bound (UCB1) algorithm \cite{UCB1}. 
While other strategies exist for the exploration-exploitation trade-off, such as $\epsilon$-greedy \cite{UCB1} and Thompson Sampling \cite{Thompson}, UCB1 was selected for its compelling theoretical guarantees and practical advantages. 
Specifically, unlike the $\epsilon$-greedy algorithm, UCB1 is parameter-free and does not require manual tuning. 
Furthermore, it features a provable logarithmic regret upper bound, ensuring efficient convergence towards the optimal policy \cite{symptoticallyoptimal}. 
Due to these favorable characteristics, this UCB1-based approach is employed. The complete procedure, which we term NaiveMAB, is detailed in Algorithm 1.
\begin{algorithm}[!t]
    \caption{NaiveMAB algorithm}
    \label{alg:AOA}
    \renewcommand{\algorithmicrequire}{\textbf{Input:}}
    \renewcommand{\algorithmicensure}{\textbf{Output:}}
    \begin{algorithmic}[1]
        \REQUIRE The set of nodes $\mathcal{N}$, the set of super arms $\boldsymbol{\mathcal{A}}$;
        \ENSURE The super arm node $i$ chooses for $(t+1)$-th transmission;
        \STATE \textbf{Initialization:} 
        \STATE $t=0$, $T^t_i(\boldsymbol{a})=0$, $\overline{R}_i^t(\boldsymbol{a})=0$, $n_i\in \mathcal{N}, \boldsymbol{a}\in \boldsymbol{\mathcal{A}}$;
        \STATE For each node $i$, try all the super arm at least once, observe $r_i^t(\boldsymbol{a})$, update $t$, $T^t_i(\boldsymbol{a})$, $\overline{R}_i^t(\boldsymbol{a})$;
        \WHILE{the network does not satisfy the stop condition}
            \STATE \textbf{Expected Reward Update (After $n_i$ completes its $t$-th transmission):}
            \STATE Observe $r_i^t(\boldsymbol{a})$ based on Eq. (\ref{naive_reward});
            \STATE Update $\overline{R}_i^t(\boldsymbol{a})\leftarrow \overline{R}_i^{t-1}(\boldsymbol{a})+\frac{1}{T_i^t(\boldsymbol{a})}[r_i^t(\boldsymbol{a})-\overline{R}_i^{t-1}(\boldsymbol{a})]$; 
            \STATE $T^t_i(\boldsymbol{a})\leftarrow T^{t-1}_i(\boldsymbol{a})+1$;
            \STATE \textbf{Super Arm Selection ($n_i$ select parameters for its $(t+1)$-th transmission):}
            \FOR{$\boldsymbol{a}\in\boldsymbol{\mathcal{A}}$}
                \STATE Compute $\widehat{R}_i^t(\boldsymbol{a})=\overline{R}_i^t(\boldsymbol{a})+c\cdot\sqrt{\frac{\log(t)}{2T_i^t(\boldsymbol{a})}}$;
            \ENDFOR
            \STATE Select super arm: $a=\underset{\boldsymbol{a}\in\boldsymbol{\mathcal{A}}}{\operatorname*{\operatorname*{argmax}}}\widehat{R}_i^t(\boldsymbol{a})$;
            \STATE Set $t=t+1$;
        \ENDWHILE
    \end{algorithmic}
\end{algorithm}

However, this conventional MAB formulation suffers from two fundamental limitations that hinder its practical application in dense LoRaWAN networks.
First, the excessive action space leads to slow convergence. 
With $|\mathcal{CF}| \cdot |\mathcal{SF}| \cdot |\mathcal{TP}|$ possible super arms, an agent must conduct an exhaustive number of trials to explore each super arm even once. 
This challenge is exacerbated in a multi-agent setting, where the network environment is non-stationary due to the interfering transmissions of other nodes. 
Consequently, the learning process converges slowly.
Second, the reward of the super arm lacks the necessary information \cite{CMABstrength}. 
The reward function in Eq. (\ref{naive_reward}) provides only a single signal for the super arm, typically reflecting the PDR. 
This approach treats the arm as a black box, ignoring the individual contributions of its constituent parameters. 
As a result, the optimization objective is limited to PDR and cannot capture other critical performance dimensions, most notably the energy consumption associated with the chosen transmission power. 
To overcome these deficiencies, this paper proposes a novel distributed resource allocation algorithm based on CMAB, termed D-LoRa.
\subsection{D-LoRa Design}
\begin{figure}[!t]
\centering
    \includegraphics[width=1\linewidth]{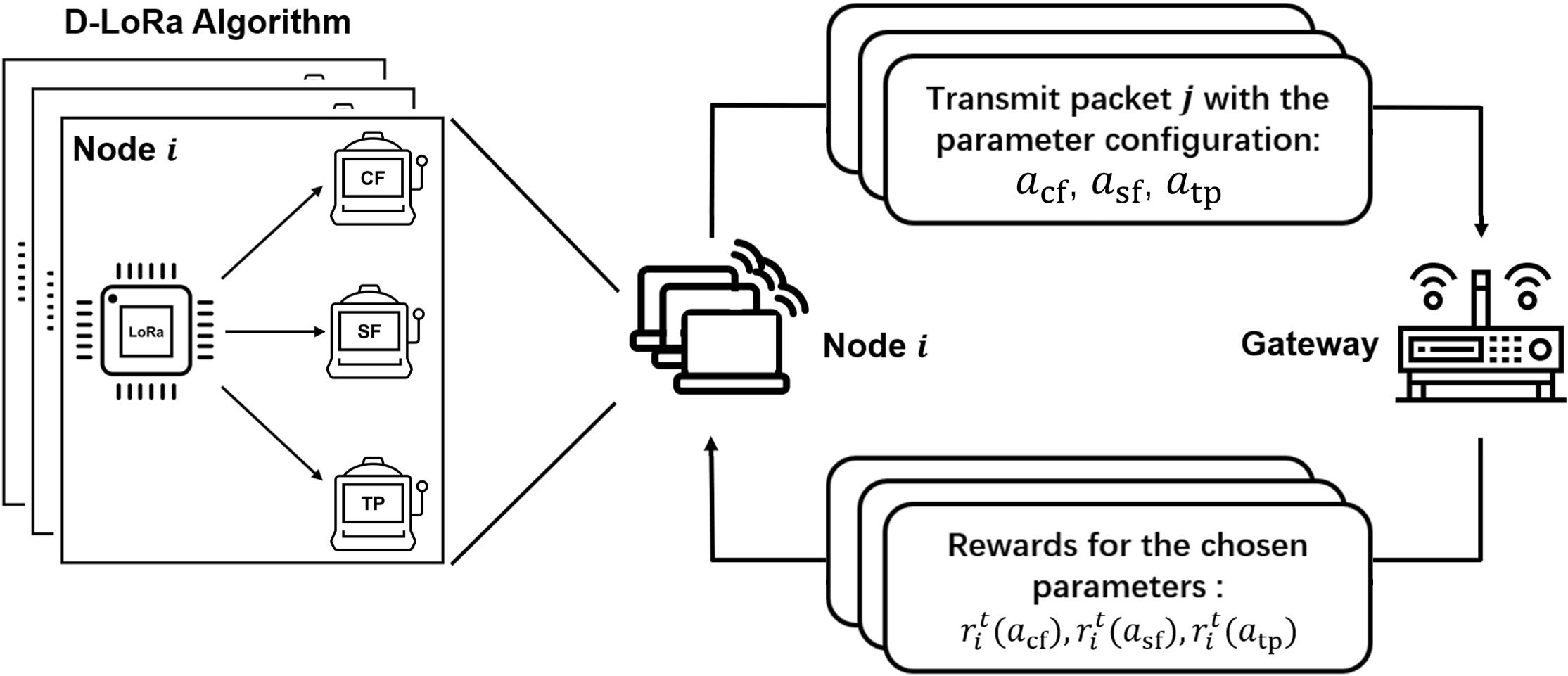}
    \caption{Framework of the D-LoRa algorithm}
    \label{D-LoRa}
\end{figure}
The proposed D-LoRa framework, depicted in Fig. \ref{D-LoRa}, redesigns the agent's learning process by moving from a simple MAB to a CMAB model. 
Instead of treating each parameter configuration as a super arm, D-LoRa decomposes the decision into selecting one base arm from each of the three parameter dimensions: CF, SF, and TP. 
The combination of these chosen base arms, (i.e., $a_{\text{cf}}, a_{\text{sf}}, a_{\text{tp}}$), then constitutes the super arm used for the actual transmission.
Critically, the feedback mechanism is also disaggregated: the agent observes distinct reward signals for each constituent base arm, rather than a single reward for the combination. 
This approach enables the agent to learn the specific contribution of each parameter choice to the transmission result. 
To facilitate this, we have designed three distinct reward functions for the base arms:

\textit{1) CF:} To prevent collisions arising from intra-channel transmissions, nodes are incentivized to select orthogonal CFs. Therefore, $ r^{t}_{i}(a_{\text{cf}})$ is formulated as:
\begin{equation}
   r^{t}_{i}(a_{\text{cf}})=\mathbb{I}\{C_j=0\land S_j=0\},
    \label{term2}
\end{equation}

\textit{2) SF:} The design of the reward function for the SF arm is driven by the dual objectives of minimizing energy consumption and ensuring link reliability. By incentivizing lower SFs, an agent reduces its packet's ToA, which conserves energy and mitigates channel contention. This pursuit of efficiency must be balanced against the need for a sufficiently robust link to guarantee successful reception. To capture this trade-off, $r^{t}_{i}(a_{\text{sf}})$ is formulated as:
\begin{equation}
  r^{t}_{i}(a_{\text{sf}})=\mathbb{I}\{C_j=0\wedge S_j=0\}+\xi\cdot\frac{\frac{\mathrm{SF}_j}{2^{\mathrm{SF}_j}}}{\sum_{\mathrm{SF}_k\in\mathcal{SF}}\frac{\mathrm{SF}_k}{2^{\mathrm{SF}_k}}},
    \label{term1}
\end{equation}
where $\xi$ is the metric factor, which controls the agent's bias towards smaller SFs. A larger value of $\xi$ increases the reward for selecting smaller SFs, thereby encouraging the agent to prioritize higher data rates and reduced ToA.

\textit{3) TP:} The selection of TP also considers the trade-off between energy efficiency and link reliability. While lower TP values significantly conserve a node's energy, they also reduce the RSSI and SINR at the gateway, increasing the risk of packet loss. The optimal strategy is therefore to use the minimum TP required to maintain a reliable communication link. Accordingly, the reward function $r^{t}_{i}(a_{\text{tp}})$ is designed as:
\begin{equation}
   r^{t}_{i}(a_{\text{tp}})=\mathbb{I}\{C_j=0\wedge S_j=0\}+\eta\cdot(1-\frac{\mathrm{TP}_j}{\sum_{\mathrm{TP}_k\in\mathcal{TP}}\mathrm{TP}_k}),
    \label{term3}
\end{equation}
where $\eta$ is the metric factor, which controls the agent's tendency towards smaller TPs. Larger $\eta$ encourages the agent to choose smaller TPs to reduce energy consumption.
\begin{algorithm}[!t]
    \caption{D-LoRa algorithm}
    \label{alg:AOA}
    \renewcommand{\algorithmicrequire}{\textbf{Input:}}
    \renewcommand{\algorithmicensure}{\textbf{Output:}}
    \begin{algorithmic}[1]
        \REQUIRE The set of nodes $\mathcal{N}$, the set of three base arms $\mathcal{CF}$, $\mathcal{SF}$,$\mathcal{TP}$;
        \ENSURE The super arm node $i$ chooses for $(t+1)$-th transmission;
        \STATE \textbf{Initialization:} 
        \STATE $t=0$, $T^t_i(a_{\text{cf}})=0$, $\overline{R}_i^t(a_{\text{cf}})=0$, $n_i\in \mathcal{N}, a_{\text{cf}}\in \mathcal{CF}$;
        \STATE $T^t_i(a_{\text{sf}})=0$, $\overline{R}_i^t(a_{\text{sf}})=0$, $n_i\in \mathcal{N}, a_{\text{sf}}\in \mathcal{SF}$;
        \STATE $T^t_i(a_{\text{tp}})=0$, $\overline{R}_i^t(a_{\text{tp}})=0$, $n_i\in \mathcal{N}, a_{\text{tp}}\in \mathcal{TP}$;
        \STATE For each node $i$, try all the base arm at least once, observe $r_i^t(a_{\text{cf}})$, $r_i^t(a_{\text{sf}})$, $r_i^t(a_{\text{tp}})$ update $t$, $T^t_i(a_{\text{cf}})$, $\overline{R}_i^t(a_{\text{cf}})$, $T^t_i(a_{\text{sf}})$, $\overline{R}_i^t(a_{\text{sf}})$, $T^t_i(a_{\text{tp}})$, $\overline{R}_i^t(a_{\text{tp}})$;
        \WHILE{the network does not satisfy the stop condition}
            \STATE \textbf{Expected Reward Update (After $n_i$ completes its $t$-th transmission):}
            \STATE Observe $r_i^t(a_{\text{cf}})$, $r_i^t(a_{\text{sf}})$, $r_i^t(a_{\text{tp}})$ based on Eq. (\ref{term2})-(\ref{term3});
            \STATE Update $\overline{R}_i^t(a_{\text{cf}})$, $\overline{R}_i^t(a_{\text{sf}})$, $\overline{R}_i^t(a_{\text{tp}})$ based on Eq. (\ref{DLoRa_expected_reward}); 
            \STATE $T^t_i(a_{\text{cf}})\leftarrow T^{t-1}_i(a_{\text{cf}})+1$, $T^t_i(a_{\text{sf}})\leftarrow T^{t-1}_i(a_{\text{sf}})+1$, $T^t_i(a_{\text{tp}})\leftarrow T^{t-1}_i(a_{\text{tp}})+1$;
            \STATE \textbf{Super Arm Selection ($n_i$ select parameters for its $(t+1)$-th transmission):}
            \FOR{$a_{\text{cf}}\in\mathcal{CF}$, $a_{\text{sf}}\in\mathcal{SF}$, $a_{\text{tp}}\in\mathcal{TP}$}
                \STATE Compute $\widehat{R}_i^t(a_{\text{cf}})$, $\widehat{R}_i^t(a_{\text{sf}})$, $\widehat{R}_i^t(a_{\text{tp}})$ based on Eq. (\ref{DLoRa_estimated_reward});
            \ENDFOR
            \STATE Select super arm based on Eq. (\ref{CUCB});
            \STATE Set $t=t+1$;
        \ENDWHILE
    \end{algorithmic}
\end{algorithm}

In D-LoRa, the learning process operates at the level of individual base arms rather than super arms.
Consequently, each agent maintains and updates a distinct expected reward, $\overline{R}(a)$, for every available base arm $a \in \mathcal{CF} \cup \mathcal{SF} \cup \mathcal{TP}$. 
Because this update mechanism is identical for all three parameter categories, we present a generalized formulation. 
The expected reward for any given base arm $a$ is updated analogously to Eq. (\ref{naive_expected_reward}):
\begin{equation}
   \overline{R}_i^t(a)=\overline{R}_i^{t-1}(a)+\frac{1}{T_i^t(a)}[r_i^t(a)-\overline{R}_i^{t-1}(a)].
    \label{DLoRa_expected_reward}
\end{equation}
To solve the formulated CMAB problem, we employ the Combinatorial UCB (CUCB) algorithm \cite{CMAB}, which adapts the UCB principle for combinatorial action spaces. 
Instead of learning the value of each super arm independently, the CUCB agent maintains an estimated reward for each base arm. 
The estimated reward for each base arm is obtained similarly to Eq. (\ref{naive_estimated_reward}):
\begin{equation}
   \widehat{R}_i^t(a)=\overline{R}_i^t(a)+c\cdot\sqrt{\frac{\log(t)}{2T_i^t(a)}}.
    \label{DLoRa_estimated_reward}
\end{equation}
The agent then chooses the super arm for the next transmission by solving for the combination of base arms that has the maximum estimated reward sum:
\begin{equation}
   \boldsymbol{a}=\underset{a_{\text{cf}}\in\mathcal{CF}, a_{\text{sf}}\in\mathcal{SF},a_{\text{sf}}\in\mathcal{TP}}{\operatorname*{\operatorname*{argmax}}}(\widehat{R}_i^t(a_{\text{cf}})+\widehat{R}_i^t(a_{\text{sf}})+\widehat{R}_i^t(a_{\text{tp}})).
    \label{CUCB}
\end{equation}

The complete D-LoRa algorithm is shown in Algorithm 2. 
In summary, the proposed D-LoRa algorithm has two principal advantages over NaiveMAB. 
First, by decomposing the super arm into base arms, it dramatically reduces the dimensionality of the action space from a combinatorial product ($|\mathcal{CF}|\cdot|\mathcal{SF}|\cdot|\mathcal{TP}|$) to a linear sum ($|\mathcal{CF}|+|\mathcal{SF}|+|\mathcal{TP}|$). 
While D-LoRa assumes that the estimated reward for a super arm can be approximated by the sum of its base arms, it achieves a substantial improvement in convergence speed, enabling the agent to converge to an effective policy far more rapidly. Second, our disaggregated reward design explicitly addresses the critical trade-off between PDR and EE in LoRaWAN networks. By assigning distinct rewards to the CF, SF, and TP arms, D-LoRa navigates this trade-off, achieving a policy that ensures reliable transmission while minimizing energy consumption.

\subsection{CD-LoRa Design}
While the fully distributed learning in D-LoRa offers high adaptability, its convergence can be slow in practice. 
The requirement for each node to learn an optimal policy across the entire joint action space of CF, SF, and TP results in a prolonged exploration phase, incurring substantial time and energy costs, which can be a bottleneck for practical deployments.

To overcome this limitation, we propose CD-LoRa, a hybrid framework that accelerates convergence by strategically decoupling the resource allocation problem. 
This decomposition is motivated by the work in \cite{central8}. 
As analysis in Section \ref{model}, the allocation of CF primarily governs inter-device, intra-channel collisions and can be addressed from a global perspective to ensure network-wide load balancing. 
In contrast, the optimal allocation of SF and TP is intricately coupled with the device-specific link quality and the complex trade-off between PDR and EE, making this subproblem ideally suited for localized, online adaptation.

Capitalizing on this insight, CD-LoRa partitions the joint optimization task. It begins with a lightweight, gateway-assisted initialization phase (i.e., CAASI). 
In CAASI, the gateway leverages its global information to assign a fixed channel to each node and prune its subsequent action space. 
With the channel allocation problem solved, each node's assigned channel remains static, allowing it to perform a more focused and rapid online learning process to optimize only the SF/TP allocation, based on the principles of D-LoRa.
The entire CAASI process can be divided into three steps, as shown in Algorithm 3:
\begin{algorithm}[!t]
    \caption{CAASI algorithm}
    \label{alg:AOA}
    \renewcommand{\algorithmicrequire}{\textbf{Input:}}
    \renewcommand{\algorithmicensure}{\textbf{Output:}}
    \begin{algorithmic}[1]
        \REQUIRE The set of nodes $\mathcal{N}$, gateway GW, the set of channels $\mathcal{CF}$;
        \ENSURE The channel assigned to node $i$ $\text{CF}_i$ and the set of valid SFs $\mathcal{SF}^*_{i}$;
        \STATE \textbf{Step 1. Data Collection:} 
        \FOR{$i=0:\left\lfloor \frac{N}{|\mathcal{CF}|}\right\rfloor$}
            \STATE Select nodes with $id$ from $|\mathcal{CF}|\cdot i$ to $|\mathcal{CF}|\cdot i + |\mathcal{CF}|-1$ to form the transmission node set $\mathcal{N}_{s}$;
            \FOR{$j=0:|\mathcal{CF}|-1$}
            \STATE Nodes in $\mathcal{N}_{s}$ choose $\mathcal{CF}[id+j\mod |\mathcal{CF}|]$ as the transmission channel;
            \STATE All nodes transmit using maximum SF and TP, GW records RSSI;
            \ENDFOR    
        \ENDFOR
        \STATE \textbf{Step 2. Channel Allocation:}
        \STATE Calculate $Q(c_{j})$ for each channel $j\in\mathcal{CF}$;
        \STATE Calculate $V(n_i)$ for each node $n_i\in\mathcal{N}$;
        \STATE Divide the nodes into $|\mathcal{CF}|$ equi-sized groups and perform the rank-based, order-preserving assignment;
        \STATE \textbf{Step 3. Action Space Initialization:} 
        \STATE Select an uninitialized node from each group of nodes to form $\mathcal{N}_{s}$;
        \FOR{$K=0:|\mathcal{SF}|-1$}
            \STATE Nodes in $\mathcal{N}_{s}$ transmit using maximum TP and $\mathcal{SF}[k]$;
            \IF{node's $\text{PDR} < \text{PDR}_{\text{min}}$}
                \STATE Delete $\mathcal{SF}[k]$ from the node's action space;
            \ENDIF    
        \ENDFOR
        \STATE Go to line 14;
    \end{algorithmic}
\end{algorithm}

\textbf{Data Collection:} The initial objective is to collect the data of the Channel State Information (CSI) for every node-to-gateway link. 
To achieve this, a coordinated data collection phase is executed. 
Transmissions are scheduled using a Time Division Multiple Access (TDMA) \cite{TDMA} approach, ensuring that transmissions are orthogonal and free from intra-channel interference. 
Specifically, nodes transmit sequentially across each available channel, with only one node active in a channel at any given time. 
To maximize the link budget and ensure successful decoding even on high-loss channels, all packets are sent using the maximum TP and the largest (most robust) SF. 
For each successful transmission, the gateway collects the RSSI data and stores the data in a node-channel matrix that maps the link quality for every possible node-channel pair.

\textbf{Channel Allocation:} The channel allocation subproblem is designed to solve a dual-objective optimization challenge: 1) enhancing link reliability for nodes with high path loss and 2) ensuring a balanced network load to mitigate intra-channel collisions. 
To address this, we propose a low-complexity, deterministic matching algorithm that systematically maps nodes to channels based on their empirical link quality.

The algorithm first establishes preference orderings for both channels and nodes. 
We define a channel quality metric, $Q(c_j)$, as the average RSSI of all packets received on each channel $j \in \mathcal{CF}$. 
A higher $Q(c_j)$ value signifies superior propagation characteristics. 
Concurrently, we define a node vulnerability metric, $V(n_i)$, as the inverse of its average RSSI, where a higher $V(n_i)$ indicates a more challenging link condition.

The core of our strategy is a rank-based, order-preserving assignment procedure. 
First, the set of channels $\mathcal{CF}$ is sorted in descending order of quality, and the set of nodes $\mathcal{N}$ is sorted in descending order of vulnerability (i.e., from weakest to strongest link). 
The sorted list of nodes is then partitioned into $|\mathcal{CF}|$ equi-sized groups. 
The assignment function, $\phi: \mathcal{N} \to \mathcal{CF}$, creates a monotonic mapping by assigning the group of nodes with the highest vulnerability (i.e., the poorest link quality) to the highest-quality channel. 
This sequential process continues, matching the group with the $k$-th highest vulnerability to the $k$-th best channel.

This principled assignment strategy has two significant benefits. 
Firstly, it provides crucial link budget compensation to the most vulnerable nodes, directly improving their PDR by allocating superior radio resources. 
Secondly, the partitioning ensures a uniform distribution of the node population across the available spectrum, preventing traffic concentration and thus fulfilling the load-balancing objective.

\textbf{Action Space Initialization:} The final step of CAASI is to accelerate the subsequent online learning phase through a priori action space pruning. 
For nodes with poor average link quality, certain lower SFs may offer insufficient link budget to ever achieve reliable communication, even at maximum transmission power. 
Allowing the learning agent to discover these infeasible SFs through trial-and-error exploration would needlessly expend time and energy, impeding convergence.
To prevent this, each node performs a one-time link feasibility test on its assigned channel. 
It transmits a sequence of packets at maximum power for each available SF.
Any SF whose resulting PDR fails to meet a minimum reliability threshold, $\text{PDR}_{\min}$, is deemed unviable for that specific node and is pruned from its action space. 
This procedure concludes by furnishing each node $i$ with a tailored and pre-validated action space, $\mathcal{SF}_i^* \subseteq \mathcal{SF}$, which contains only demonstrably effective SFs. 
This significantly reduces the exploration burden on the subsequent distributed learning algorithm, leading to faster policy convergence.

\begin{algorithm}[!t]
    \caption{CD-LoRa algorithm}
    \label{alg:AOA}
    \renewcommand{\algorithmicrequire}{\textbf{Input:}}
    \renewcommand{\algorithmicensure}{\textbf{Output:}}
    \begin{algorithmic}[1]
        \REQUIRE The set of nodes $\mathcal{N}$, the set of three base arms $\mathcal{CF}$, $\mathcal{SF}$,$\mathcal{TP}$;
        \ENSURE The super arm node $i$ chooses for $(t+1)$-th transmission;
        \STATE \textbf{Initialization:}
        \STATE Allocate channel and initialize action space according to Algorithm 3;
        \STATE $t=0$, $T^t_i(a_{\text{sf}})=0$, $\overline{R}_i^t(a_{\text{sf}})=0$, $n_i\in \mathcal{N}, a_{\text{sf}}\in \mathcal{SF}$;
        \STATE $T^t_i(a_{\text{tp}})=0$, $\overline{R}_i^t(a_{\text{tp}})=0$, $n_i\in \mathcal{N}, a_{\text{tp}}\in \mathcal{TP}$;
        \STATE For each node $i$, try all the base arm at least once, observe $r_i^t(a_{\text{sf}})$, $r_i^t(a_{\text{tp}})$ update $t$,  $T^t_i(a_{\text{sf}})$, $\overline{R}_i^t(a_{\text{sf}})$, $T^t_i(a_{\text{tp}})$, $\overline{R}_i^t(a_{\text{tp}})$; 
        \WHILE{the network does not satisfy the stop condition}
            \STATE \textbf{Expected Reward Update (After $n_i$ completes its $t$-th transmission):}
            \STATE Observe $r_i^t(a_{\text{sf}})$, $r_i^t(a_{\text{tp}})$ based on Eq. (\ref{term1})-(\ref{term3});
            \STATE Update $\overline{R}_i^t(a_{\text{sf}})$, $\overline{R}_i^t(a_{\text{tp}})$ based on Eq. (\ref{DLoRa_expected_reward}); 
            \STATE $T^t_i(a_{\text{sf}})\leftarrow T^{t-1}_i(a_{\text{sf}})+1$, $T^t_i(a_{\text{tp}})\leftarrow T^{t-1}_i(a_{\text{tp}})+1$;
            \STATE \textbf{Super Arm Selection ($n_i$ select parameters for its $(t+1)$-th transmission):}
            \FOR{$a_{\text{sf}}\in\mathcal{SF}_i^*$, $a_{\text{tp}}\in\mathcal{TP}$}
                \STATE Compute $\widehat{R}_i^t(a_{\text{sf}})$, $\widehat{R}_i^t(a_{\text{tp}})$ based on Eq. (\ref{DLoRa_estimated_reward});
            \ENDFOR
            \STATE Select super arm: $\boldsymbol{a}=\underset{\boldsymbol{a}\in\boldsymbol{\mathcal{A}}}{\operatorname*{\operatorname*{argmax}}}(\widehat{R}_i^t(a_{\text{sf}})+\widehat{R}_i^t(a_{\text{tp}}))$, $\boldsymbol{a}=\{a_{\text{sf}}$, $a_{\text{tp}}\}$;
            \STATE Set $t=t+1$
        \ENDWHILE
    \end{algorithmic}
\end{algorithm}
Upon completion of the centralized CAASI procedure, the network transitions to its operational phase of distributed learning. 
Building upon the D-LoRa framework, each node independently learns a policy to dynamically select the optimal SF and TP. 
The learning problem in CD-LoRa is significantly less complex than D-LoRa.
First, the permanent channel allocation reduces the agent's task from a three-dimensional (CF/SF/TP) to a two-dimensional (SF/TP) problem. 
Second, the action space initialization conducted during CAASI may have further reduced the number of available SF arms for each node. 
This substantial reduction in the dimensionality of the learning space enables the agent to converge to an optimal policy more rapidly. 
The complete CD-LoRa algorithm, which integrates the CAASI phase with this distributed learning, is detailed in Algorithm 4.
\subsection{Property Analysis}
In terms of computational complexity, NaiveMAB incurs the highest computational cost at $O(N \cdot |\mathcal{CF}| \cdot |\mathcal{SF}| \cdot |\mathcal{TP}|)$ due to its exhaustive evaluation of the entire super arm space. In contrast, D-LoRa substantially reduces this complexity to $O(N \cdot (|\mathcal{CF}| + |\mathcal{SF}| + |\mathcal{TP}|))$ by decomposing the problem into independent base arms. Building on this, CD-LoRa achieves the lowest online complexity of $O(N \cdot (|\mathcal{SF}| + |\mathcal{TP}|))$. This efficiency is gained by offloading the channel allocation to the one-time centralized initialization phase (with a setup complexity of $O(N\cdot(|\mathcal{CF}| + |\mathcal{SF}|))$.

The regret $\mathcal{R}(T)$ over $t$ transmissions is defined as the expected difference between the cumulative reward of the optimal strategy and the algorithm’s cumulative reward:
\begin{equation}
    \mathcal{R}(t)=t\cdot r^*-\mathbb{E}\left[\sum_{\tau=1}^t r_{\tau}\right],
    \label{regret}
\end{equation}
where $r^*$ is the expected reward of the optimal arm and $r_{\tau}$ is the reward obtained at $\tau$-th transmission.

\textit{Theorem 1}. (Regret Upper Bound) The regret of NaiveMAB, D-LoRa, CD-LoRa is bounded by
\begin{equation}
    \mathcal{R}(T)\leq O(\sqrt{Kt\ln t}),
    \label{upper_Bound}
\end{equation}
where $K$ is the number of super arms for NaiveMAB and the number of base arms for D-LoRa and CD-LoRa.

\textit{Proof:}  See \cite{CMAB,UCB1} for proofs.

\textit{Corollary 1}. (Asymptotically Optimality) D-LoRa is asymptotically optimal.

\textit{Proof:}  According to \textit{Theorem 1}, there exists a constant $C$ such that, for sufficiently large $t$:
\begin{equation}
    \mathcal{R}(t)\leq C\sqrt{Kt\ln t},
    \label{upper_Bound_constant}
\end{equation}
To prove asymptotic optimality, we need to examine the average regret per round:
\begin{equation}
    \frac{\mathcal{R}(t)}{t}\leq\frac{C\sqrt{Kt\ln t}}{t}=C\sqrt{\frac{K\ln t}{t}}.
    \label{average_regret}
\end{equation}
Evaluate the limit as $t\to\infty$:
\begin{equation}
    \lim_{t\to\infty}\frac{\mathcal{R}(t)}{t}\leq\lim_{t\to\infty}C\sqrt{\frac{K\ln t}{t}}=0.
    \label{limit}
\end{equation}
According to Eq. (\ref{regret}), $\mathcal{R}(t)>0$. Therefore, $\lim_{t\to\infty}\frac{\mathcal{R}(t)}{t}=0$ and the asymptotically optimality is proved.

\textit{Corollary 2}. (Convergence Speed) The convergence speeds of the three algorithms satisfy CD-LoRa $>$ D-LoRa $>$ NaiveMAB.

\textit{Proof:} According to the proof of \textit{Corollary 1}, Eq. (\ref{average_regret}) implies that smaller $K$ leads to faster convergence speed.

%% file: sections/5-simulation.tex
\section{Simulation Experiments}
\label{simulation}
\subsection{Simulation Setup}
The simulations are carried out on our LoRaWAN simulator LoRaSimPlus \cite{Simulator}. We consider a LoRaWAN network with nodes randomly distributed in a circular area. The available LoRa parameters for each node are set as: $\mathcal{SF}=\{7,8,9,10,11,12\}$, $\mathcal{CF}=\{868.1, 868.3, 868.5, 868.7, 868.9, 869.1, 869.3, 869.5\}$ MHz, $\mathcal{TP}=\{2,4,6,8,10,12,14\}$ dBm. The packet generation interval satisfies the Poisson distribution. The parameters of the Log-Distance Path Loss Model are set according to \cite{PathLoss}, and the detailed parameter settings are shown in Table \ref{Parameter_Setting}.
\begin{table}[htbp]
\caption{Parameter setting}
\centering
\renewcommand{\arraystretch}{1.2}
\begin{tabular}{c|c}
\hline
\textbf{Parameters}&\textbf{Values}\\
\hline
Number of gateway& 1\\
Packet payload size (PS)&50 bytes\\
Average packet generation interval&20 s\\
Bandwidth (BW)&125 kHz\\
Code Rate (CR)&4/5\\
Reference distance ($d_0$)&1000 m\\
Average path loss ($\overline{L^{\mathrm{pl}}}(d_0)$)& 128.95 dB\\
Path loss factor ($\gamma$)&1\\
Standard deviation of shadowing effect ($\delta_{1}$)&7.8\\
Standard deviation of AWGN ($\delta_{2}$)&1\\
Weight Factor ($c$)&2\\
Metric Factors ($\xi$, $\eta$)& 1, 1.8\\
PDR threshold ($\text{PDR}_{\min}$)& 25 \%\\
\hline
\end{tabular}
\label{Parameter_Setting}
\end{table}

We conducted various simulation experiments in both stationary and nonstationary environments to compare the performance of the aforementioned algorithms (NaiveMAB, D-LoRa, CD-LoRa) against the random algorithm and two state-of-the-art resource allocation algorithms:
\begin{itemize}
\item \textbf{Random}: Each node selects LoRa parameters randomly for packet transmission. 
\item \textbf{CAASI+ADR} \cite{central1}: ADR is a centralized algorithm controlled by a network server that dynamically adjusts the data transmission rate of devices based on signal quality and network conditions. Since the original ADR algorithm does not provide CF allocation, we combine CAASI with ADR for CF/SF/TP allocation.
\item \textbf{MIX-MAB} \cite{distributed4}: MIX-MAB is a distributed algorithm that treats each parameter configuration as an arm and combines EXP3 and SE algorithm to find the best arm.
\end{itemize}
\subsection{Simulation Results in Stationary Environment}
The stationary environment means stationary channel conditions: all channels in $\mathcal{CF}$ use the same path loss parameters shown in Table \ref{Parameter_Setting} and they do not change over time. 

\subsubsection{\textit{Performance Comparison as the Number of Nodes Changes}} 
\begin{figure}[!t] 
  \centering
  \begin{subfigure}[t]{0.9\columnwidth} 
    \centering
    \includegraphics[width=\linewidth]{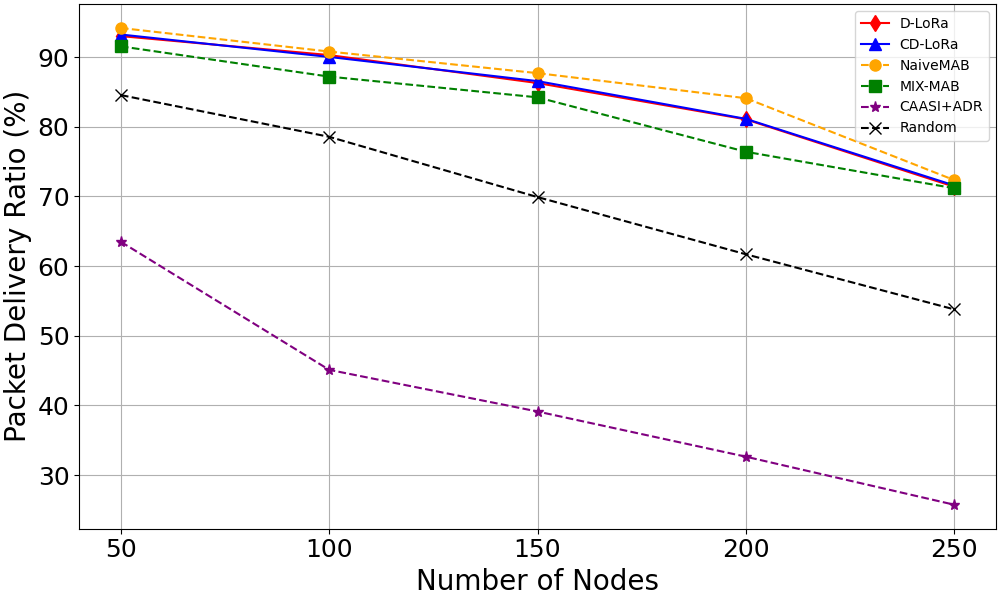}
    \caption{}
    \label{PDR_Num_Homo}
  \end{subfigure}

  \begin{subfigure}[t]{0.9\columnwidth} 
    \centering
    \includegraphics[width=\linewidth]{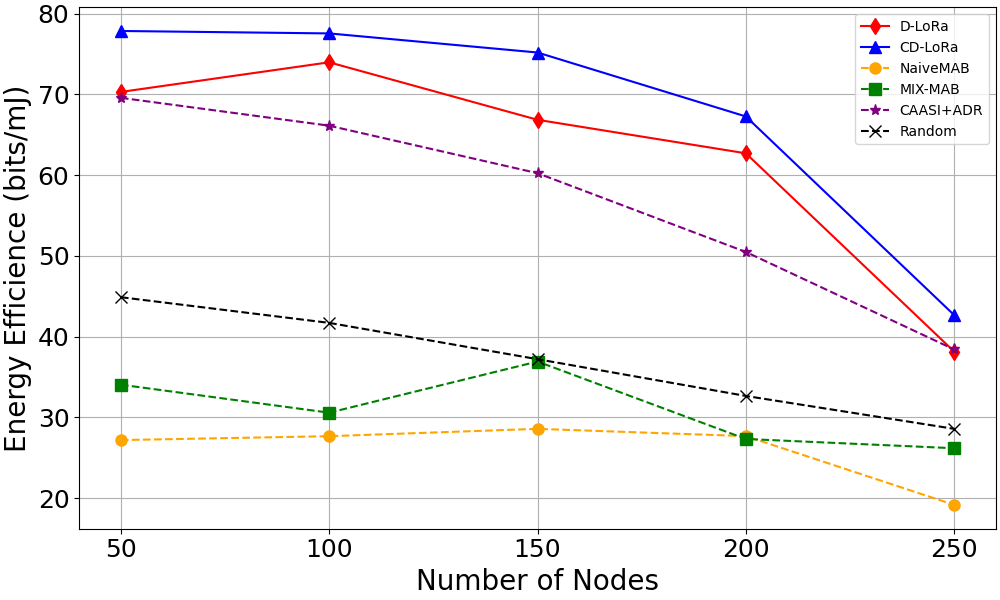}
    \caption{} 
    \label{EE_Num_Homo}
  \end{subfigure}
  
  \caption{Performance comparison of the algorithms as the number of nodes changes in the stationary environment: (a) PDR comparison, (b) EE comparison.}
  \label{Perform_Num}
\end{figure}

First, we examine the effect of increasing the number of nodes on algorithm performance when the network topology radius is fixed at 1000 m. 
The performance of the six algorithms with the number of nodes increasing from 50 to 250 is shown in Fig. \ref{Perform_Num}. 
From the experimental results, it can be observed that as the number of nodes in the network increases, the performance of the algorithms decreases.
This performance degradation is attributed to the heightened probability of packet collisions. Specifically, an increase in the number of nodes leads to more frequent concurrent transmissions, which consequently elevates the collision rate and degrades both the PDR and EE.

\begin{figure}[!t]
    \centering
    
    \begin{subfigure}[b]{0.48\linewidth}
        \centering
        \includegraphics[width=\linewidth]{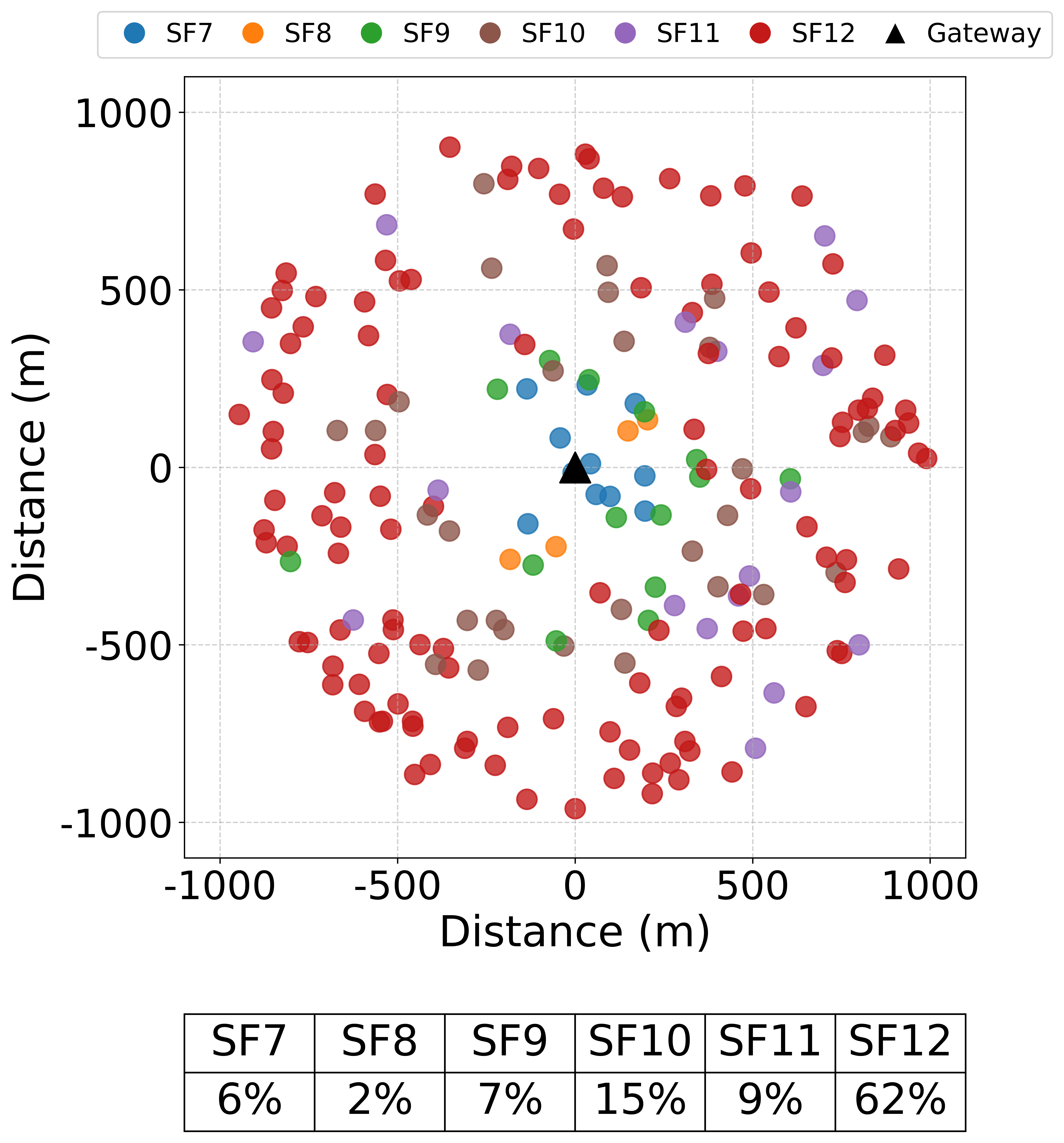}
        \caption{}
        \label{ADR_SF_Distribution} 
    \end{subfigure}
    \hfill 
    \begin{subfigure}[b]{0.48\linewidth}
        \centering
        \includegraphics[width=\linewidth]{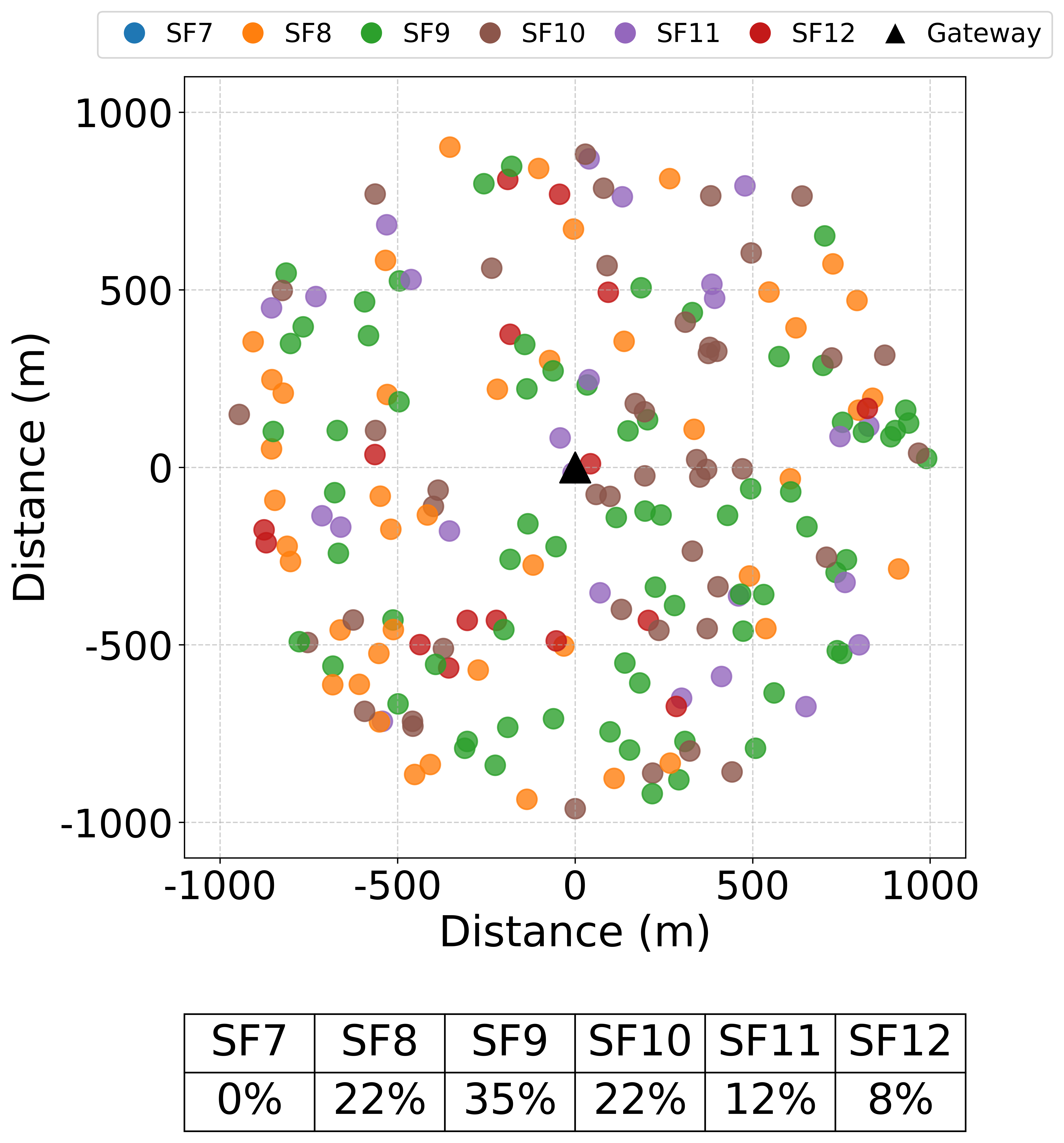}
        \caption{}
        \label{DLoRa_SF_Distribution}
    \end{subfigure}
    
    \caption{Comparison of SF distributions in a network of 200 nodes in the stationary environment: (a) ADR, (b) D-LoRa.}
    \label{SF_distributions}
\end{figure}

As illustrated in Fig. \ref{Perform_Num}(a), all four online learning algorithms, D-LoRa, CD-LoRa, NaiveMAB, and MIX-MAB, outperform the Random and CAASI+ADR benchmarks in terms of PDR. While the PDR of all algorithms degrades with an increasing number of nodes, the proposed learning-based approaches exhibit greater resilience. For instance, as the number of nodes increases from 50 to 250, the PDR of D-LoRa declines by only 21.6\% in contrast to the 37.7\% drop experienced by CAASI+ADR.
The severe PDR degradation of the ADR mechanism comes from its resource allocation strategy. 
ADR tends to assign the same SF to nodes located at similar distances from the gateway. 
This approach leads to severe intra-SF collisions, especially in dense networks. The resulting SF distribution is visualized in Fig. \ref{SF_distributions} for a network of 200 nodes. As depicted in Fig. \ref{SF_distributions}(a), ADR allocates SF12 to a staggering 62\% of the nodes, creating a major collision bottleneck. In contrast, Fig. \ref{SF_distributions}(b) demonstrates that D-LoRa promotes a more balanced SF distribution, where nodes autonomously select diverse SFs based on historical feedback to mitigate collisions.

Fig. \ref{Perform_Num}(b) presents the EE comparison. The results show that CD-LoRa and D-LoRa consistently and substantially outperform all other algorithms across the entire range of network densities.
Specifically, in a sparse network with 50 nodes, CD-LoRa achieves the highest EE of 78 bits/mJ, closely followed by D-LoRa at 70 bits/mJ. 
These values are approximately 2.9 and 2.6 times higher than that of NaiveMAB (27 bits/mJ) and 2.3 and 2.1 times higher than that of MIX-MAB (34 bits/mJ), respectively.
This superior performance is maintained even as the network scales, underscoring the robustness of our proposed mechanisms in optimizing energy consumption.


\subsubsection{\textit{Performance Comparison as the Network Topology Radius Changes}} 
\begin{figure}[!t]
  \centering
  \begin{subfigure}[t]{0.9\columnwidth} 
    \centering
    \includegraphics[width=\linewidth]{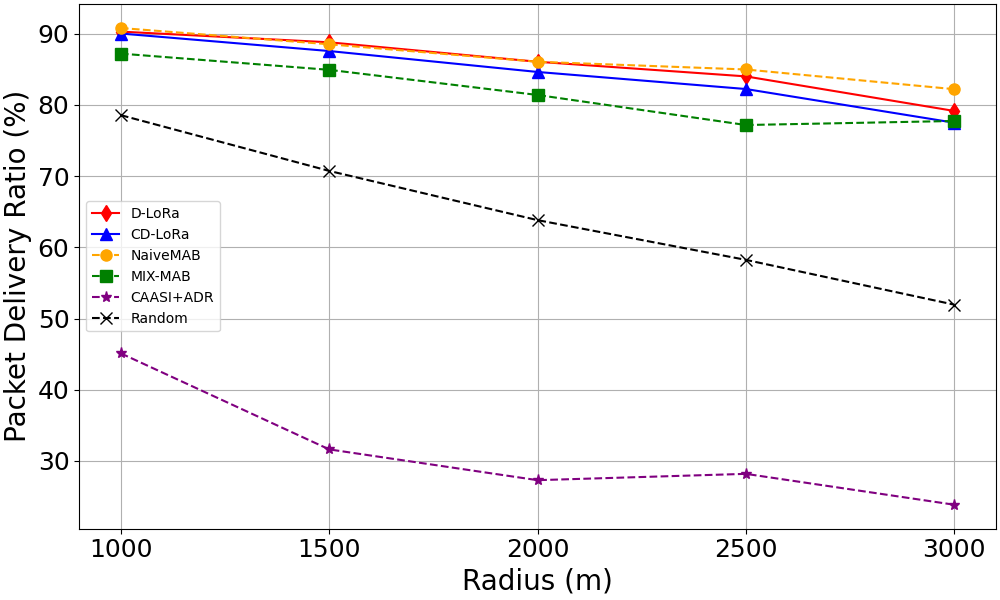}
    \caption{}
    \label{PDR_Radius_Homo}
  \end{subfigure}

  \begin{subfigure}[t]{0.9\columnwidth} 
    \centering
    \includegraphics[width=\linewidth]{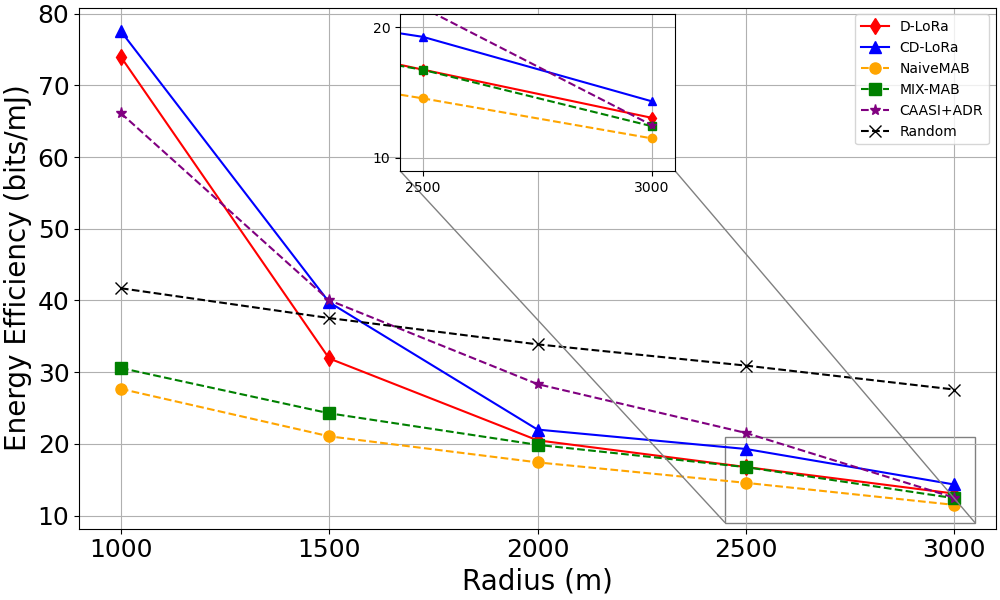}
    \caption{}
    \label{EE_Radius_Homo}
  \end{subfigure}
  
  \caption{Experimental results of the algorithms as the network radius changes in the stationary environment: (a) PDR comparison, (b) EE comparison.}
  \label{both_radius_homo}
\end{figure}

We further investigate the impact of the network topology radius on the performance of the algorithms under a fixed network density of 100 nodes. 
Fig. \ref{both_radius_homo} illustrates the PDR and EE, respectively, as a function of the network radius, which is varied from 1000 m to 3000 m.
The observed performance degradation with increasing network radius is a direct consequence of greater signal path loss. 
Larger distances between nodes and the gateway necessitate the use of higher TP and larger SF to ensure successful reception. 
While these adjustments aim to counteract signal attenuation, they inherently reduce energy efficiency due to longer ToA and increased power consumption. 
Furthermore, the prolonged ToA heightens the likelihood of packet collisions, which in turn diminishes the PDR.

Fig. \ref{both_radius_homo}(a) illustrates the PDR performance as a function of the network radius. 
A clear performance gap is observed between the learning-based approaches and the benchmarks. 
The four online learning algorithms (D-LoRa, CD-LoRa, NaiveMAB, and MIX-MAB) consistently outperform both Random and CAASI+ADR. 
Moreover, the learning-based algorithms demonstrate greater robustness to the expanding network coverage. 
As the radius increases from 1000 m to 3000 m, the PDR for these algorithms declines by less than 15\%, whereas Random suffers a much steeper drop of 26.6\%. 
Among the learning algorithms, NaiveMAB achieves the highest PDR, followed closely by D-LoRa and CD-LoRa, all of which show a clear advantage over MIX-MAB.

\begin{figure}[!t]
    \centering
    \includegraphics[width=1\linewidth]{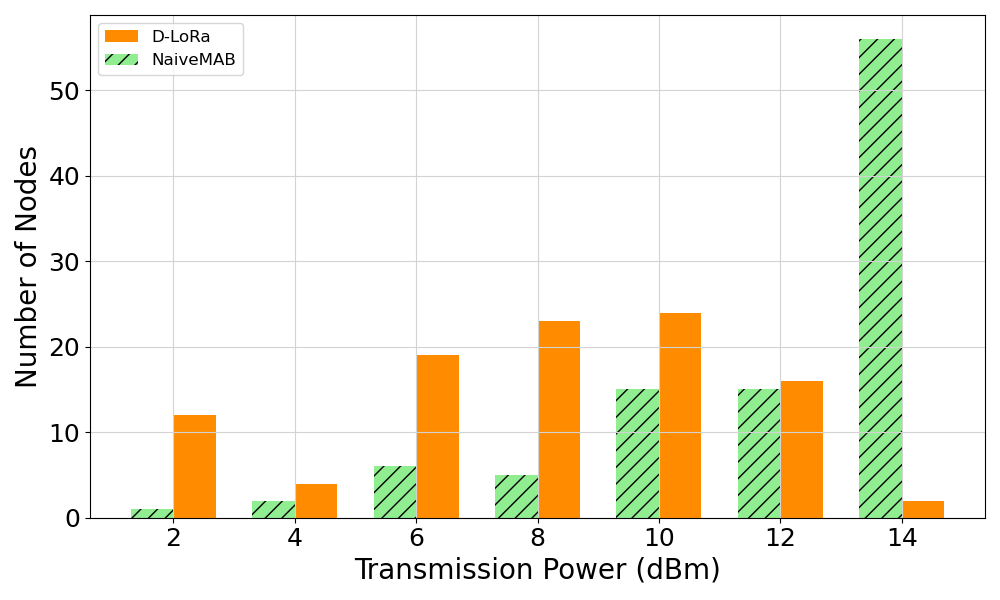}
    \caption{Comparison of TP usage between D-LoRa and NaiveMAB in the network with 100 nodes and 1000m radius in the stationary environment.}
    \label{TP_allocation}
\end{figure}

The EE results, presented in Fig. \ref{both_radius_homo}(b), reveal a nuanced performance trade-off that is dependent on the network radius.
In smaller-radius deployments (e.g., 1000 m), CD-LoRa and D-LoRa exhibit superior EE, achieving 78 bits/mJ and 74 bits/mJ, respectively. 
This performance is substantially higher than that of MIX-MAB (31 bits/mJ) and NaiveMAB (27 bits/mJ). 
This advantage is attributed to the adaptive power control mechanism embedded in the reward function of D-LoRa and CD-LoRa. 
Specifically, these algorithms incentivize nodes near the gateway to select lower TP to conserve energy, a behavior that MIX-MAB and NaiveMAB lack, as they tend to use the higher TP to maximize reception probability regardless of distance. 
This contrast in TP strategy is further substantiated by Fig. \ref{TP_allocation}, which shows that over 50\% of nodes under NaiveMAB use the maximum TP, whereas D-LoRa promotes the use of lower TP to minimize redundant energy expenditure.
Conversely, as the network radius expands, a significant drop in the EE of D-LoRa and CD-LoRa is observed, to the point where their performance may fall below that of the Random algorithm. 
This phenomenon does not signify a failure, but rather illustrates a deliberate design trade-off. 
To maintain a high PDR for distant nodes, our algorithms strategically allocate larger SFs and higher TPs to counteract severe path loss. 
While this strategy successfully keeps the PDR high (as seen in Fig. \ref{both_radius_homo}(a)), it comes at the inevitable cost of increased energy consumption, thereby reducing EE. 
This highlights a fundamental trade-off between ensuring reliability for nodes and optimizing overall network energy efficiency.

\subsubsection{\textit{Convergence Comparison}}
\begin{figure}[!t] 
  \centering
  \begin{subfigure}[t]{0.9\columnwidth} 
    \centering
    \includegraphics[width=\linewidth]{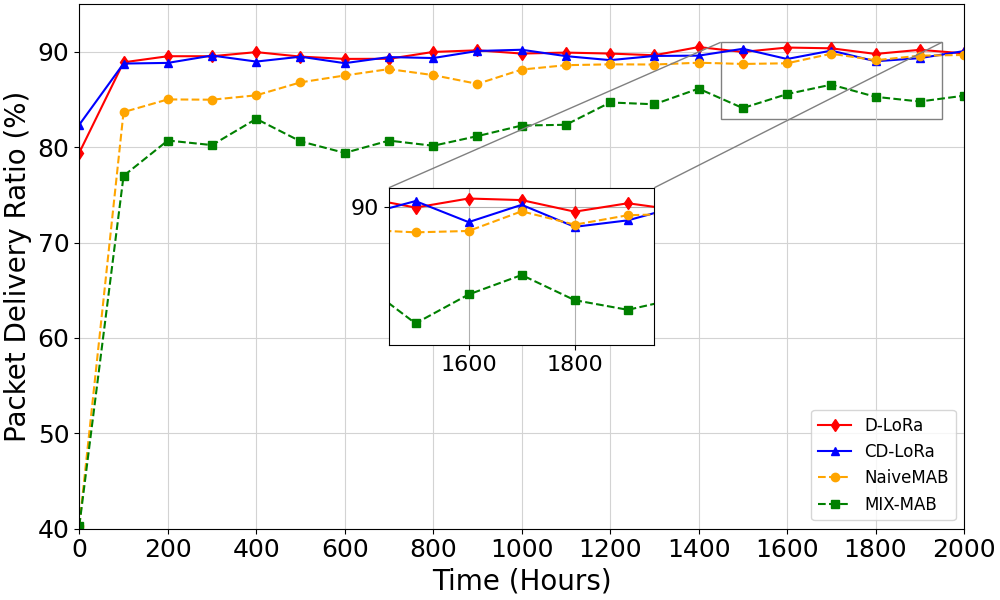}
    \caption{}
    \label{PDR_training_stationary}
  \end{subfigure}

  \begin{subfigure}[t]{0.9\columnwidth} 
    \centering
    \includegraphics[width=\linewidth]{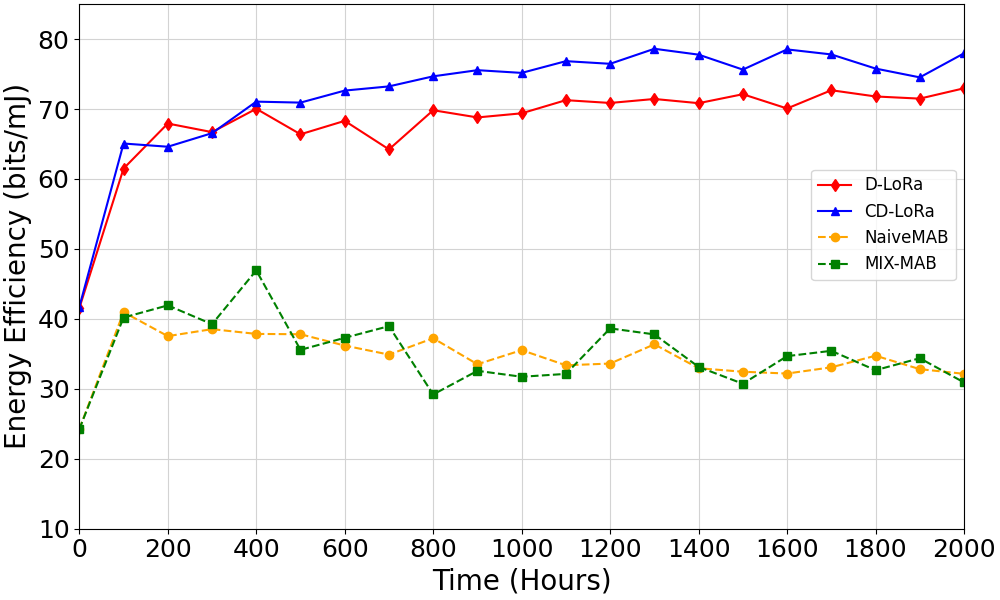}
    \caption{}
    \label{EE_training_stationary}
  \end{subfigure}
  
  \caption{Performance changes with time of the online learning algorithms in the stationary environment: (a) PDR, (b) EE.}
  \label{both_training_stationary}
\end{figure}

We now evaluate the convergence behavior of the four online learning algorithms. Fig. \ref{both_training_stationary} illustrate the evolution of PDR and EE over a 2000-hour training period in a stationary network of 100 nodes with a 1000 m radius. 
The results clearly indicate that D-LoRa and CD-LoRa exhibit significantly faster convergence and achieve superior steady-state performance compared to both NaiveMAB and MIX-MAB.

Specifically, D-LoRa and CD-LoRa demonstrate rapid initial learning, reaching approximately 90\% PDR within the first 400 hours before stabilizing in the 88-90\% range. 
Their EE shows a similar swift improvement, climbing from an initial 30 bits/mJ to a stable state above 70 bits/mJ with minimal fluctuations. 
In contrast, NaiveMAB requires around 1200 hours for its PDR to reach a comparable level, while its EE plateaus at a mere 35 bits/mJ by the end of the simulation. 
The convergence of MIX-MAB is even more sluggish, with its PDR stabilizing at a lower range of 80-85\% after 1600 hours and its EE showing negligible improvement, hovering around 30 bits/mJ amidst considerable volatility.

These empirical results underscore the efficacy of the CMAB framework. By decomposing the super arm into base arms, CMAB drastically reduces the number of arms each agent must explore. 
This structure is the primary reason for the rapid convergence observed in D-LoRa and CD-LoRa. 
Furthermore, a fine-grained comparison between our two proposed algorithms reveals that CD-LoRa converges slightly faster and achieves a higher EE. 
This suggests that in a stationary environment, the centralized CAASI (as implemented in CD-LoRa) can more efficiently identify the optimal resource allocation policy by leveraging a global view, thereby validating its suitability for such scenarios.

\subsection{Simulation Results in Nonstationary Environment}
To evaluate the adaptability of the proposed algorithms, we now shift our focus to another common scenario: non-stationary environment. This environment is characterized by two key features: 1) heterogeneous channel conditions, where each of the eight channels exhibits distinct attenuation properties governed by different path loss model parameters, and 2) dynamic channel state transitions, where the properties of all channels change abruptly at a specific point during the simulation. The detailed path loss parameters at the reference distance, $\overline{L^{\mathrm{pl}}}(d_0)$, for each channel before and after this abrupt change are specified in TABLE \ref{Channel_Condition_before} and TABLE \ref{Channel_Condition_after}, respectively.
\begin{table}[htbp]
\caption{$\overline{L^{\mathrm{pl}}}(d_0)$ (dB) of the channels before the change}
\renewcommand{\arraystretch}{1.2}
\centering
\begin{tabular}{c|cccccccc}
\hline
Channel &CF1&CF2&CF3&CF4&CF5&CF6&CF7&CF8 \\
\hline
$\overline{L^{\mathrm{pl}}}(d_0)$&136&134&132&130&128&126&124&122 \\
\hline
\end{tabular}
\label{Channel_Condition_before}
\end{table}
\begin{table}[htbp]
\caption{$\overline{L^{\mathrm{pl}}}(d_0)$ (dB) of the channels after the change}
\renewcommand{\arraystretch}{1.2}
\centering
\begin{tabular}{c|cccccccc}
\hline
Channel &CF1&CF2&CF3&CF4&CF5&CF6&CF7&CF8 \\
\hline
$\overline{L^{\mathrm{pl}}}(d_0)$&122&124&126&128&130&132&134&136 \\
\hline
\end{tabular}
\label{Channel_Condition_after}
\end{table}

This experimental setup was deliberately engineered to represent a worst-case scenario, featuring a complete inversion of channel quality. Specifically, channels that were initially favorable (i.e., having low path loss) were abruptly switched to an unfavorable state (high path loss), and vice versa. This challenging design allows for a stringent evaluation of an algorithm's adaptability and its ability to recover from drastic environmental shifts.
\begin{figure*}[!t]
  \centering
  \begin{subfigure}[t]{0.48\textwidth}
    \centering
    \includegraphics[width=\linewidth]{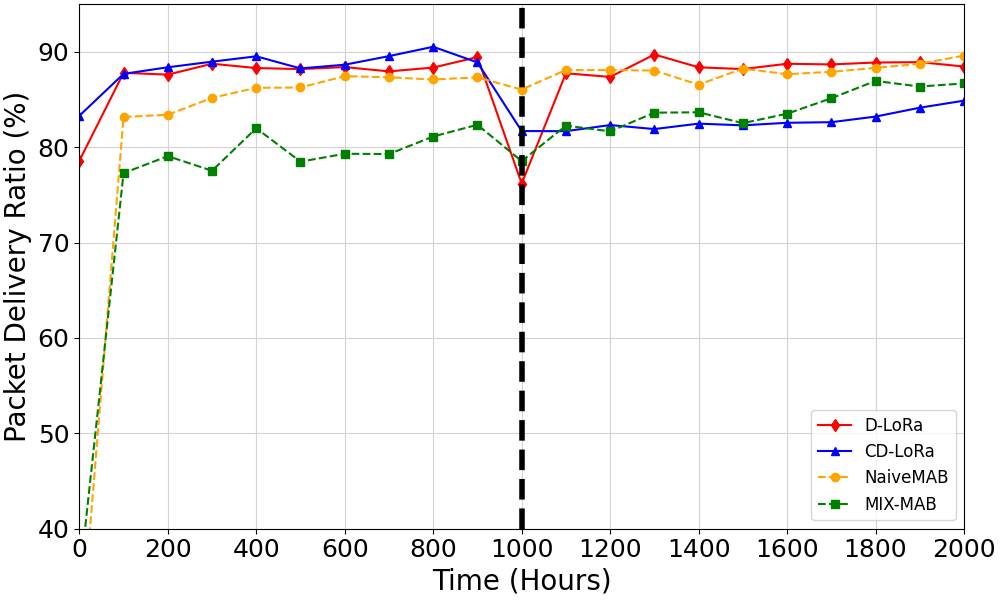}
    \caption{}
    \label{PDR_Num_Homo}
  \end{subfigure}
  \hfill
  \begin{subfigure}[t]{0.48\textwidth}
    \centering
    \includegraphics[width=\linewidth]{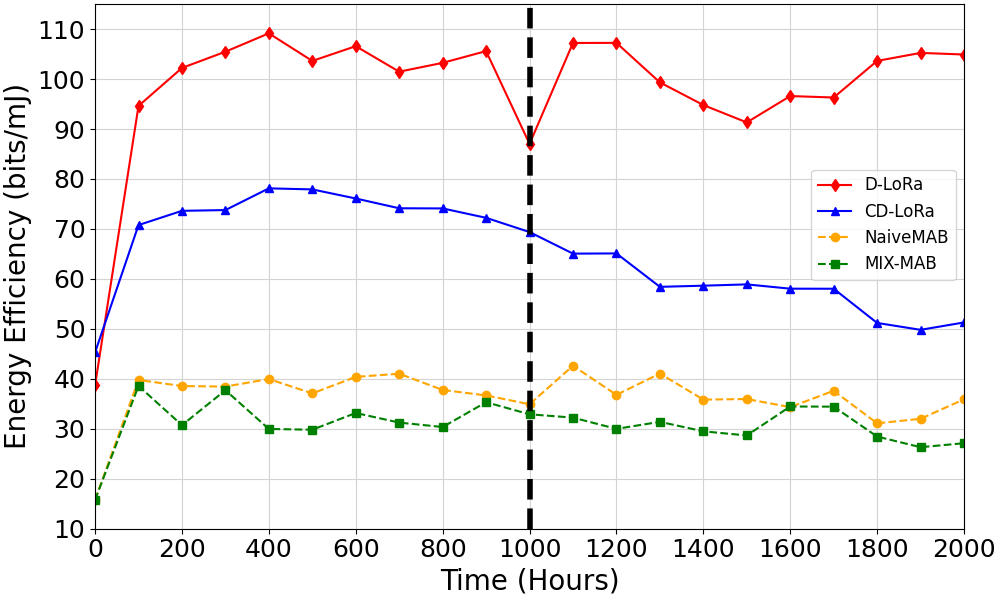}
    \caption{}
    \label{EE_Num_Homo}
  \end{subfigure}
  
  \caption{Performance changes with time of the online learning algorithms in the nonstationary environment: (a) PDR, (b) EE.}
  \label{training_process_dynamic}
\end{figure*}
\begin{figure*}[h!]
  \centering
  \begin{subfigure}[t]{0.48\textwidth}
    \centering
    \includegraphics[width=\linewidth]{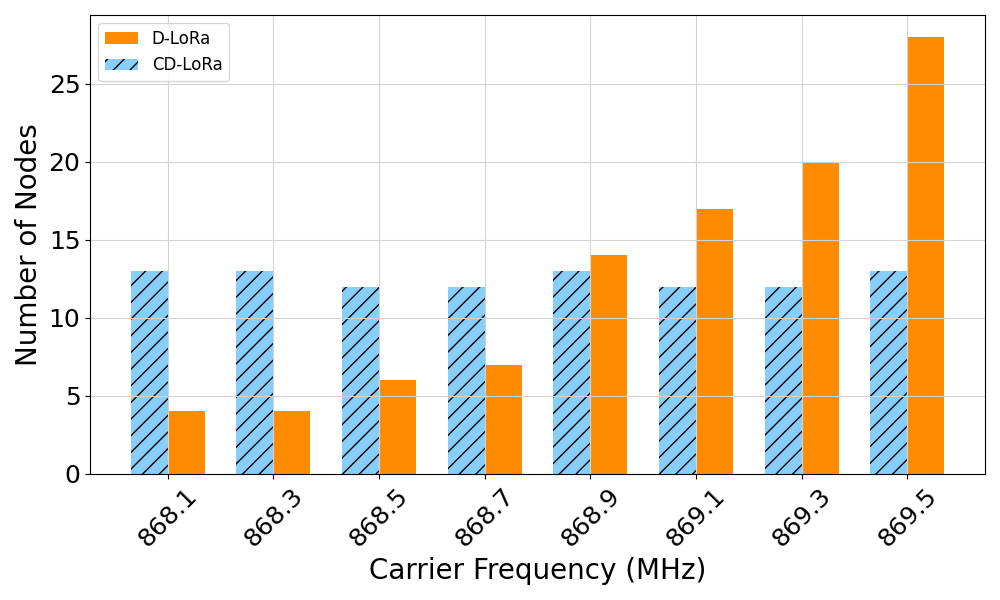}
    \caption{}
    \label{CF_allocation_at_1000h}
  \end{subfigure}
  \hfill
  \begin{subfigure}[t]{0.48\textwidth}
    \centering
    \includegraphics[width=\linewidth]{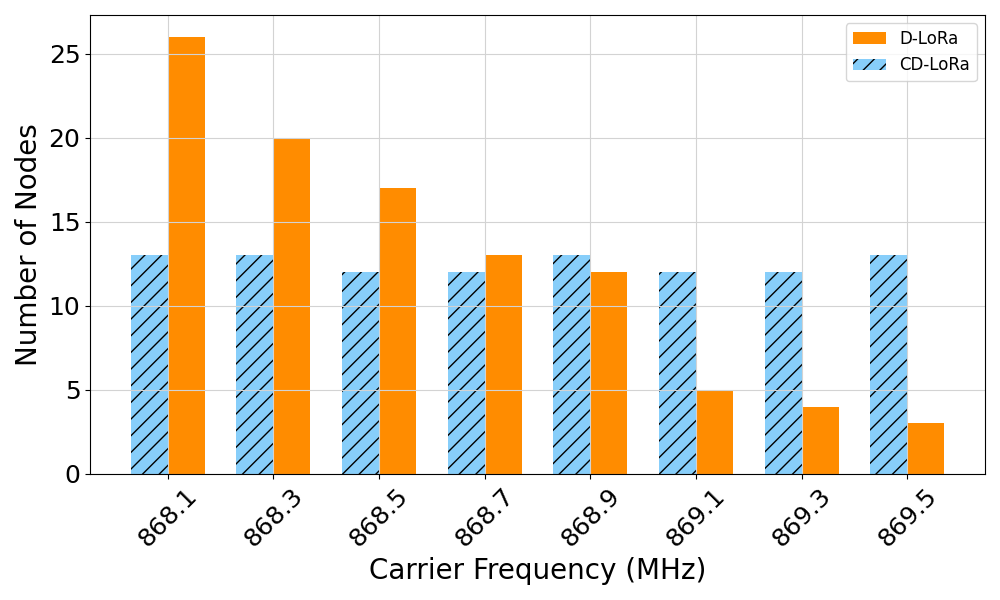}
    \caption{}
    \label{CF_allocation_at_2000h}
  \end{subfigure}
  
  \caption{Comparison of CF usage between D-LoRa and CD-LoRa before and after brute environmental change: (a) TP usage at 1000 hours, (b) TP usage at 2000 hours.}
  \label{TP_usage}
\end{figure*}
Fig. \ref{training_process_dynamic} depict the PDR and EE evolution for the different algorithms over a 2000-hour training period within a nonstationary network of 100 nodes and a 1000m radius. 
The abrupt environmental change was introduced at the 1000-hour mark. 
The results highlight the exceptional dynamic adaptability of the D-LoRa algorithm, which stands in sharp contrast to the performance degradation exhibited by CD-LoRa in nonstationary environments. 
During the initial 1000 hours, D-LoRa rapidly converges to a high-performance state, achieving a PDR of 88-90\% and an impressive EE of approximately 105 bits/mJ. 
Following the channel change, D-LoRa experienced a transient performance drop, with PDR falling to 76\%, but demonstrated remarkable resilience by rapidly recovering to the 85-90\% PDR range within just 200 hours, with its EE returning to around 105 bits/mJ. 
Conversely, CD-LoRa suffered a permanent performance degrade, failing to show any meaningful recovery and with its PDR and EE decreasing to a level below its stationary environment performance as shown in Fig. \ref{training_process_dynamic}(b).

This divergence in performance comes from the algorithms' fundamental framework differences. 
The failure of CD-LoRa can be attributed to its centralized CF allocation mechanism, which establishes a static channel allocation strategy before the learning phase begins. 
While effective in the stationary environment, this design makes it incapable of adapting its strategy when the channel conditions change. 
In contrast, the purely distributed nature of D-LoRa enables each agent to continuously reassess and adapt its strategy based on its historical information. 
This inherent architectural flexibility allows the network as a whole to fluidly respond to and recover from drastic environmental changes, making D-LoRa intrinsically suitable for nonstationary scenarios.

This behavioral difference is visually verified by the channel utilization statistics presented in Fig. \ref{TP_usage}. It is evident that D-LoRa dynamically shifts its channel allocation to track the favorable channels with low path loss. After the channel qualities are inverted at the 1000-hour mark, D-LoRa agents actively migrate their transmissions to the newly favorable channels. Significantly, the strategy is more nuanced than simply overloading the best channels; D-LoRa demonstrates an intelligent trade-off, sometimes assigning nodes closer to the gateway to less favorable channels, presumably to mitigate intra-channel interference and optimize overall network performance. CD-LoRa, bound by its one-time static assignment, is unable to make any such adjustments, continuing to utilize channels that have become severely suboptimal.

%% file: sections/6-field.tex
\section{Field Experiments}
\label{field}
\subsection{Field Experiment Setup}
\begin{figure}[!t]
    \centering
    \begin{subfigure}[b]{0.48\linewidth}
        \centering
        \includegraphics[height=0.9\linewidth]{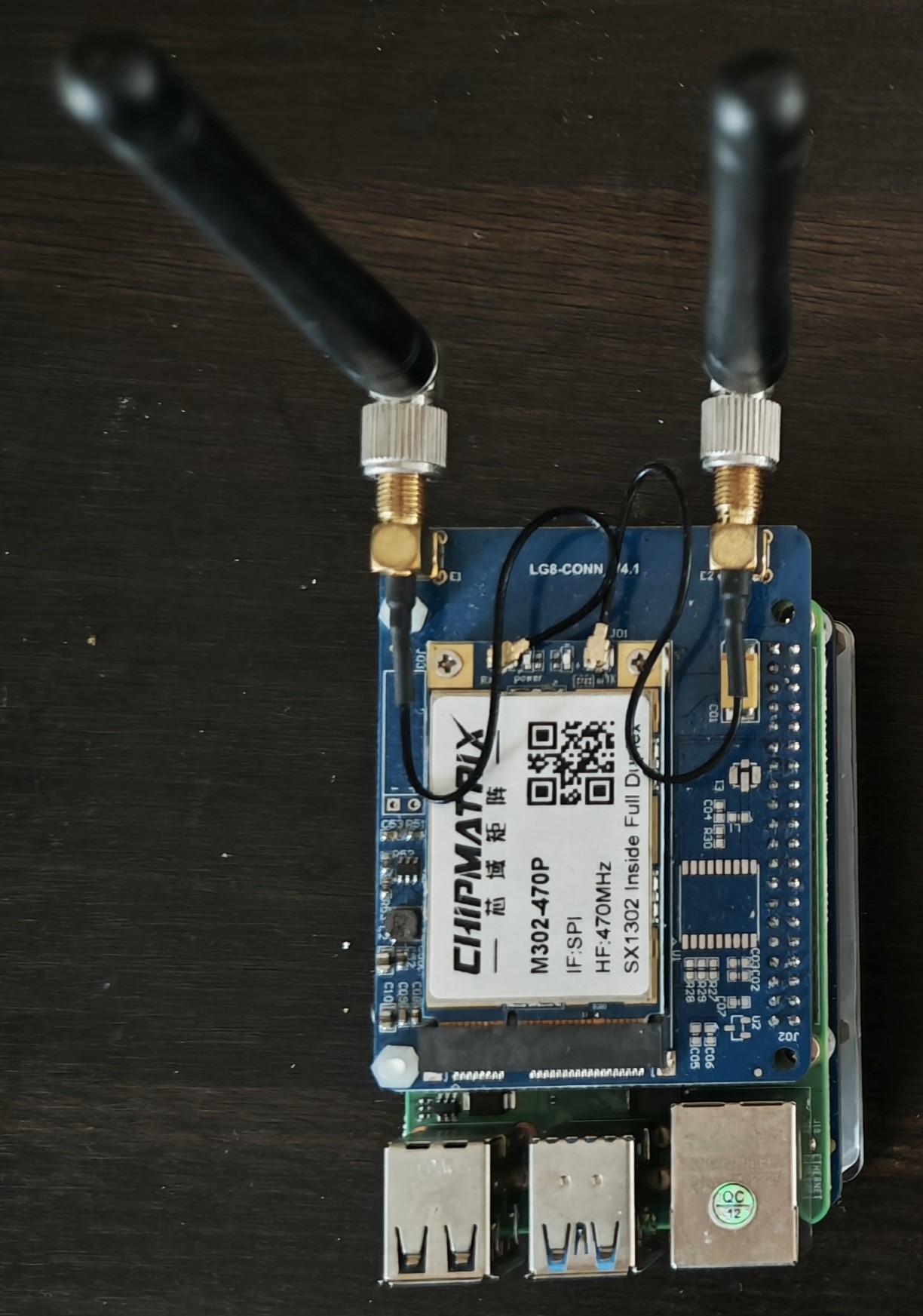}
        \caption{}
        \label{Gateway}
    \end{subfigure}
    \hfill
    \begin{subfigure}[b]{0.48\linewidth}
        \centering
        \includegraphics[height=0.9\linewidth]{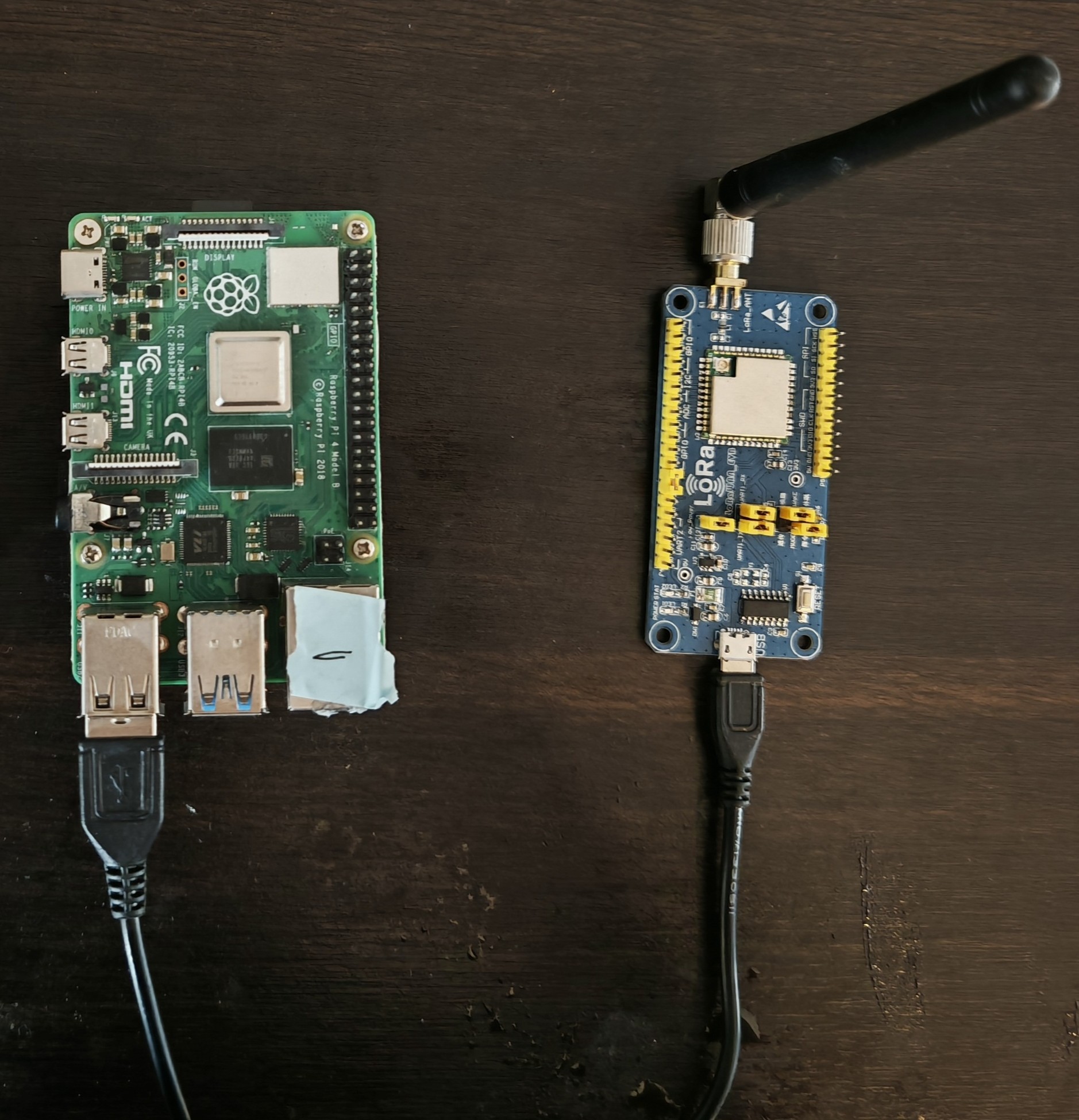}
        \caption{}
        \label{Node}
    \end{subfigure}
    \caption{Hardware equipment used in field experiments: (a) Gateway, (b) Node.}
    \label{hardware}
\end{figure}
\begin{figure}[!t]
    \centering
    \includegraphics[width=0.9\linewidth]{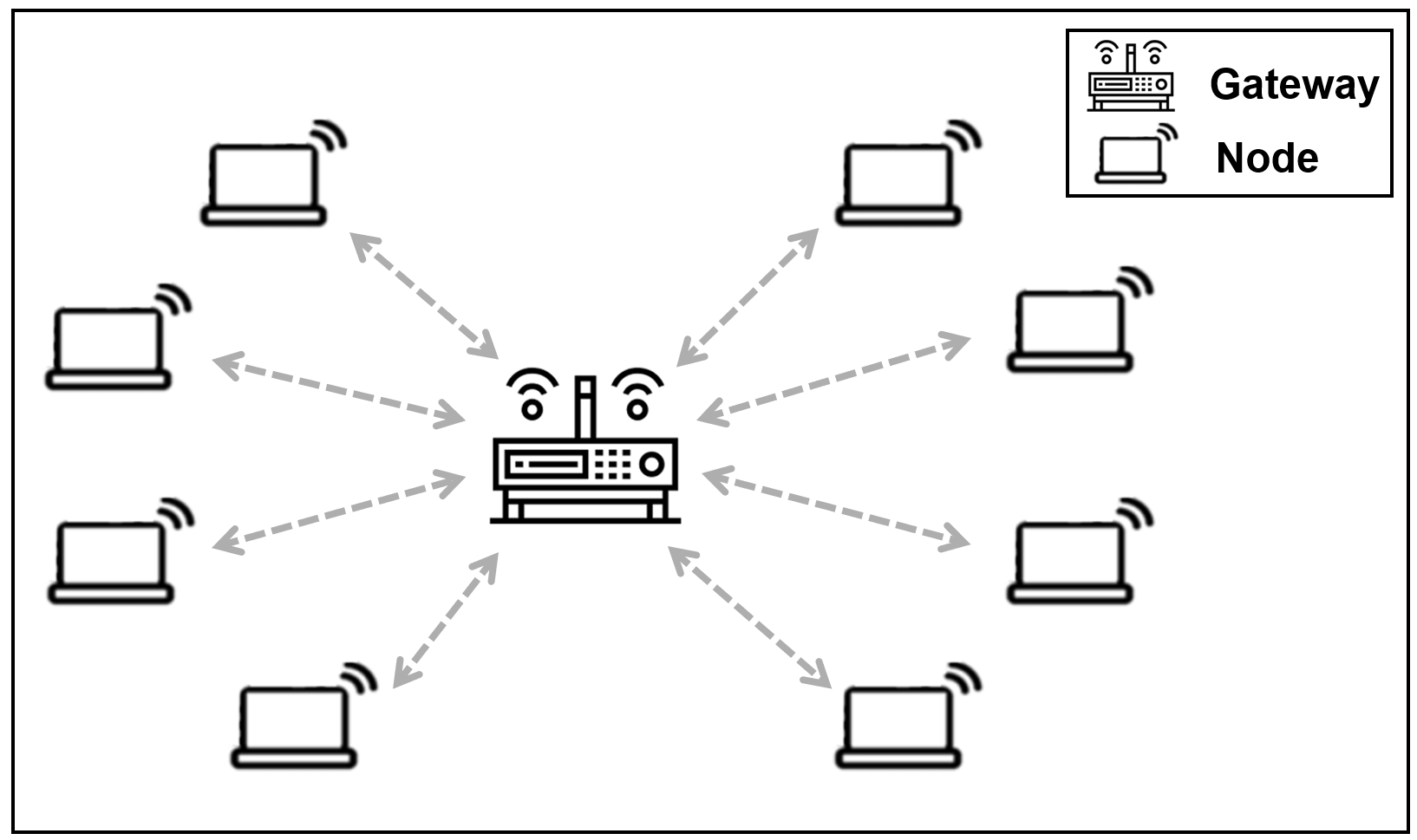}
    \caption{Network topology in the field experiment}
    \label{topology}
\end{figure}
To validate the real-world performance of our proposed algorithms, we deployed a LoRaWAN network testbed comprising eight nodes and a gateway. 
The hardware equipment used for this experiment is depicted in Fig. \ref{hardware}. Each node combines a Raspberry Pi 4B for computation with an STM32WLE5 board for radio communication, which is powered by a 5V UPS battery.
The resource allocation algorithm executes on the Raspberry Pi, which then relays the selected parameters to the STM32WLE5 via a serial interface. 
The STM32WLE5 board, equipped with a Semtech SX1278 transceiver, subsequently configures and transmits the LoRa packets \cite{STM32WLE5}. The gateway is constructed from a Raspberry Pi 4B and a Semtech SX1302 concentrator module \cite{SX1302}, enabling the simultaneous multi-channel, multi-SF reception. The testbed was deployed in a laboratory environment to simulate a dense network, with nodes placed 2–3 m away from the gateway, as shown in Fig. \ref{topology}. 
The obstacles between the nodes and the gateway are different, which lead to different link quality of different nodes.
To accommodate the on-node processing overhead required for parameter configuration, the average packet transmission interval was set to 30s. 
All other settings were configured as specified in Table \ref{Parameter_Setting}.

For the EE analysis, it is essential to clarify our methodology. 
The energy consumption metric in this study is defined to exclusively account for the energy expended during the transmission of LoRa packets. 
We acknowledge the existence of static energy overhead from ancillary processes, most notably the operation of the Raspberry Pi. 
However, since this baseline overhead is constant across all experimental runs and common to all evaluated algorithms, it does not affect the validity of the relative performance comparisons. 
Therefore, our simplified model provides a fair and robust basis for assessing the comparative efficiency of the different strategies.

\subsection{Field Experimental Results}
\begin{figure}[!t] 
  \centering
  \begin{subfigure}[t]{0.9\columnwidth} 
    \centering
    \includegraphics[width=\linewidth]{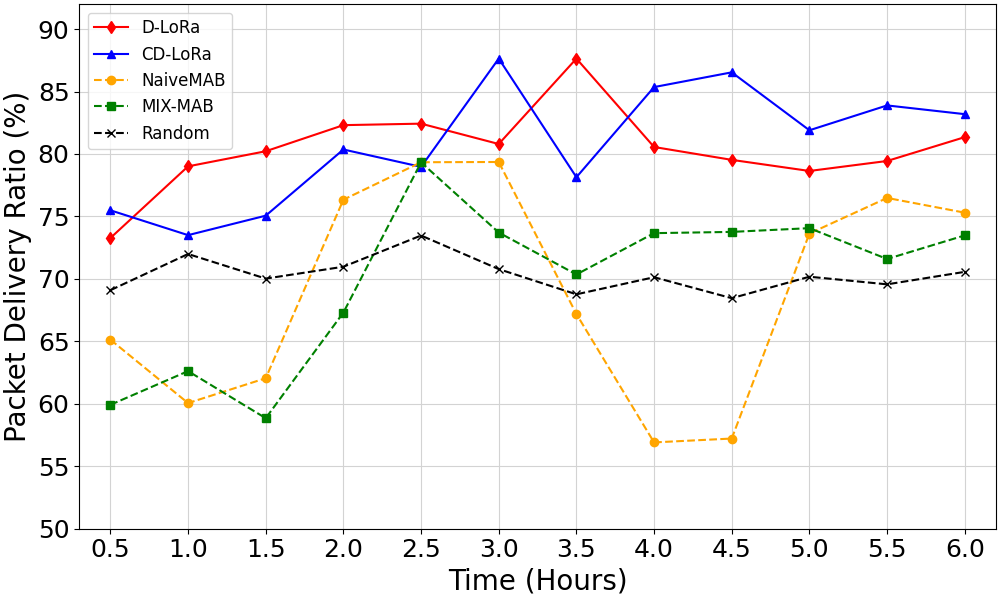}
    \caption{} 
    \label{PDR_Field}
  \end{subfigure}
  
  \begin{subfigure}[t]{0.9\columnwidth} 
    \centering
    \includegraphics[width=\linewidth]{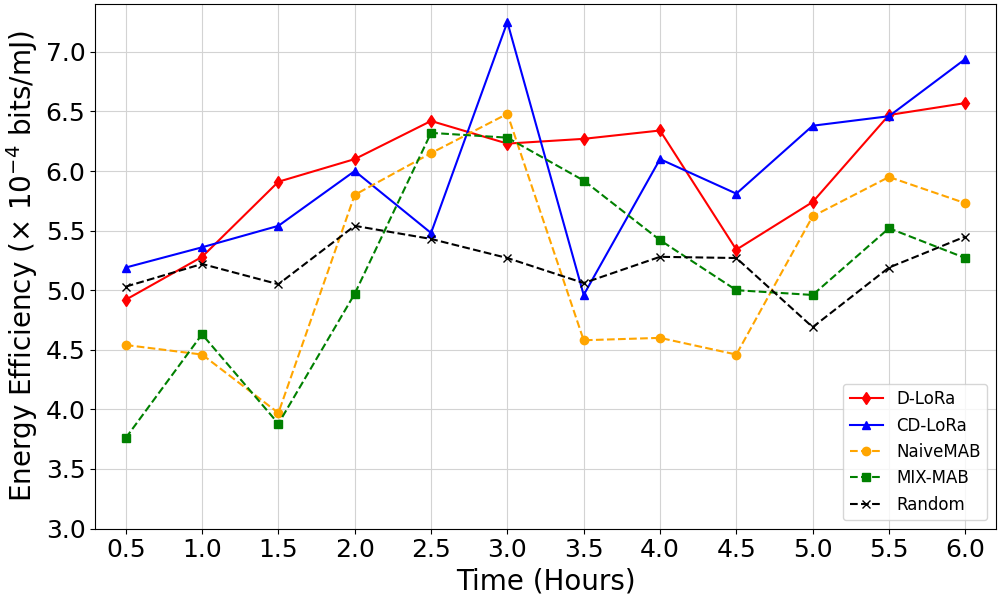}
    \caption{} 
    \label{EE_Field}
  \end{subfigure}
  \caption{Performance changes with time in the field experiment: (a) PDR, (b) EE.}
  \label{Field_Experiment}
\end{figure}
The six-hour field experiment demonstrates the clear superiority of our proposed algorithms, as illustrated by the PDR evolution in Fig. \ref{Field_Experiment}(a). 
In the final phase of the experiment (5.5–6h), D-LoRa achieved a PDR of 81.3\%, significantly outperforming the Random (70.3\%), MIX-MAB (73.4\%), and NaiveMAB (75.3\%) baselines by 15.6\%, 10.8\%, and 8.0\%, respectively. 
Notably, CD-LoRa emerged as the top performer, consistently maintaining the highest PDR after the four-hour mark and ultimately reaching 83.2\%. 
This represents a further 2.3\% improvement over D-LoRa, validating our hypothesis that the centralized CAASI phase establishes a more effective foundation for the subsequent distributed learning process.
Beyond the final performance figures, the results underscore the adaptive learning capabilities inherent in our frameworks. 
Both D-LoRa and CD-LoRa exhibited a distinct upward PDR trend throughout the experiment, indicating successful online optimization. 
In conclusion, D-LoRa's PDR improved by 10.9\% from its initial value, while CD-LoRa's increased by 10.2\%. 
In contrast, the PDR of the Random baseline remained volatile, fluctuating within a narrow 5\% range. 
This divergence strongly validates the efficacy of our online learning approach, enabling nodes to autonomously adapt and enhance network reliability in a dynamic, real-world environment.

The superior EE of our proposed algorithms is also evident in Fig. \ref{Field_Experiment}(b). During the final experimental phase (5.5–6h), D-LoRa demonstrated substantial gains, outperforming the Random, MIX-MAB, and NaiveMAB baselines by 26.1\%, 24.7\%, and 14.6\%, respectively. 
CD-LoRa further enhanced this efficiency, improving upon D-LoRa's EE by an additional 5.8\%. 
Critically, the EE for both frameworks showed a consistent upward trend, with D-LoRa's final EE increasing by 33.5\% and CD-LoRa's by 33.9\% relative to their initial values. 
These results confirm that our specialized reward function, coupled with the online learning framework, successfully optimizes for both network reliability and energy conservation.

It is worth noting that in both Fig. \ref{Field_Experiment}(a) and Fig. \ref{Field_Experiment}(b), the PDR and EE curves for MIX-MAB and NaiveMAB exhibit a significant initial decline before recovering. 
This phenomenon is a direct result of the initialization phase of the algorithms. 
This phase required them to explore every possible action before the main learning process. 
In contrast, D-LoRa and CD-LoRa leverage a CMAB-based action decomposition strategy, which obviates the need for such a prolonged initialization phase, enabling our algorithms to begin learning and improving performance rapidly from the outset.
\begin{table}[htbp]
\caption{Experimental data over the entire experiment process}
\centering
\begin{tabular}{|c|c|c|c|c|}
\hline
\textbf{\textbf{Algorithm}}&\textbf{Sent}&\textbf{Received}&\textbf{PDR}(\%)&\textbf{EE}($\times10^{-4}$ bits/mJ) \\
\hline
$\textbf{\text{CD-LoRa}}$& 5355 & 4330 & 80.86 & 5.66\\
\hline
$\textbf{\text{D-LoRa}}$& 5315 & 4275 & 80.43 & 5.94\\
\hline
$\text{NaiveMAB}$& 5294 & 3707 & 70.02 & 5.13\\
\hline
$\text{MIX-MAB}$& 5280 & 3646 & 69.05 & 5.16\\
\hline
$\text{Random}$& 5271 & 3704 & 70.27 & 5.20\\
\hline
\end{tabular}
\label{field_results}
\end{table}

TABLE \ref{field_results} presents the aggregate performance data for all algorithms over the entire experiment, including the number of packets sent, the number of packets received, PDR and EE. 
In terms of network reliability, CD-LoRa and D-LoRa achieved average PDR of 80.86\% and 80.43\%, respectively. This represents a substantial improvement of over 10\% compared to the best-performing baseline, Random (70.27\%), underscoring our algorithms' effectiveness in link reliability improvement. 
A similar advantage was observed in EE, where D-LoRa ($5.94\times 10^{-4}$ bits/mJ) and CD-LoRa ($5.66\times10^{-4}$ bits/mJ) secured the top two rankings, significantly outperforming all baselines. Collectively, these findings validate that our learning-based resource allocation algorithms successfully co-optimize for network reliability and energy conservation through intelligent parameter selection.

%% file: sections/7-conclusion.tex
\section{conclusion}
\label{cons}
In this paper, we addressed the complex challenge of joint resource allocation in LoRaWAN networks to co-optimize for PDR and EE. Our fully distributed D-LoRa framework leverages a CMAB model with disaggregated rewards to ensure high adaptability in dynamic environments, while its hybrid counterpart, CD-LoRa, integrates a lightweight, gateway-assisted initialization phase to achieve accelerated convergence in stationary conditions. Validation through extensive simulations and real-world field experiments confirms the efficacy of our approach, demonstrating that CD-LoRa excels in stable networks while D-LoRa's adaptability is superior under non-stationary channel dynamics. Critically, our proposed algorithms deliver substantial performance gains over state-of-the-art baselines, improving PDR by up to 10.8\% and EE by 26.1\% in physical deployments. Ultimately, this work validates that our online learning based resource allocation framework effectively co-optimizes for network reliability and energy conservation, offering a robust and practical solution for scalable LoRaWAN deployments.